\documentclass[%
 reprint,
 amsmath,amssymb,
 aps,
 prl,
 twocolumn,
 bibnotes
]{revtex4-2}
\usepackage{graphicx}
\usepackage{dcolumn}
\usepackage{bm}
\usepackage{float}
\usepackage{supertabular}
\usepackage{makecell}
\usepackage{booktabs}
\usepackage{comment}
\usepackage{color}
\usepackage{import}
\usepackage{multirow}
\usepackage[colorlinks,citecolor=blue]{hyperref}
\begin{document}

\title{Experimental Measurement-Device-Independent Quantum Cryptographic Conferencing}
\author{$\text{Yifeng Du}^1$}
\author{$\text{Yufeng Liu}^1$}
\author{$\text{Chengdong Yang}^1$}
\author{$\text{Xiaodong Zheng}^1$}
\author{$\text{Shining Zhu}^1$}
\author{$\text{Xiao-song Ma}^{1,2,3,}$}
\email{xiaosong.ma@nju.edu.cn}
\affiliation{$^1$National Laboratory of Solid-state Microstructures, School of Physics, College of Engineering and Applied Sciences, Collaborative Innovation Center of Advanced Microstructures, Jiangsu Physical Science Research Center, Nanjing University, Nanjing 210093,China\\
$^2$Synergetic Innovation Center of Quantum Information and Quantum Physics, University of Science and
Technology of China, Hefei, 230026, China\\
$^3$Hefei National Laboratory, Hefei 230088, China
}

\begin{abstract}
Quantum cryptographic conferencing (QCC) allows sharing secret keys among multiple distant users and plays a crucial role in quantum networks. Because of the fragility and low generation rate of genuine multipartite entangled states required in QCC, realizing and extending QCC with the entanglement-based protocol is challenging. Measurement-device-independent (MDI) QCC, which removes all detector side channels, is a feasible long-distance quantum communication scheme to practically generate multipartite correlation with multiphoton projection measurement. Here we experimentally realize the three-user MDIQCC protocol with four-intensity decoy-state method, in which we employ the polarization encoding and the Greenberger-Horne-Zeilinger state projection measurement. Our work demonstrates the experimental feasibility of the MDI QCC, which lays the foundation for the future realization of quantum networks with multipartite communication tasks.
\end{abstract}

\maketitle
\section{Introduction}
Quantum key distribution (QKD) \cite{BB84,qkd_review_4,qkd_review_1,qkd_review_2,qkd_review_3,qkd_review_5} enables two  users to share secure keys based on the principles of quantum physics with information-theoretic unconditional security. Over the past few decades, tremendous advances have been made in the research of QKD theory \cite{decoy1,decoy2,decoy3,diqkd,Pirandola2012,mdi_qkd,tf_qkd,mp_qkd_2,mp_qkd_1} and experimental techniques \cite{1120km_qkd,chen_integrated_2021,110M_qkd,1000km_qkd}, paving the way for practical applications.To facilitate a broader range of application scenarios, it is essential to expand the scope of point-to-point QKD to encompass multiple users and to develop the quantum networks \cite{qnet_review_2,qnet_review_1}. Recently, many QKD networks have been established \cite{qkd_network_2,qkd_network_3,qkd_network_1}. Going beyond the point-to-point two-user quantum communication, some multiuser communication protocols, including quantum cryptographic conferencing (QCC) \cite{qcc_concept_1,qcc_concept_2,qcc_review} and quantum secret sharing (QSS) \cite{qss_1,qss_2,qss_3,qss_4,qss_5,qss_6,qss_7,qss_8}, are capable of establishing multipartite quantum correlation among distant users. These protocols illustrate the fascinating application prospects in multiuser quantum communication.

Quantum cryptographic conferencing, as an important protocol of quantum communication, aims for sharing the cryptographic keys among multiple users to conduct secure conference. It has been proved that the global multipartite private state, from which the secure conference keys can be distilled, should be the genuine multipartite entanglement state \cite{qcc_gme_2,qcc_gme_3,qcc_gme_1} including the Greenberger-Horne-Zeilinger (GHZ) state \cite{ghz_state} and the $W$ state \cite{ghz_w_state}. There are two main categories of QCC protocols: entanglement-based protocols \cite{qcc_eb_theory_2,qcc_eb_theory_1} and time-reversal protocols \cite{pol_qcc,w_qcc_1,pm_qcc,cow_qcc_2,cow_qcc_1,cow_qcc_3,w_qcc_2,lu2024repeaterlike,xie2024multifield}. The entanglement-based protocols, which rely on the distribution of multipartite entanglement states, have been experimentally demonstrated with advanced entangled photon sources  \cite{qcc_eb_exp_1,qcc_eb_exp_2}. 
Because of the fragility and low generation rate of multipartite entanglement states, it is challenging for entanglement-based protocols to be practically implemented over long distance and among multiple users.

The multipartite correlation can also be obtained by projection measurement in time-reversal protocols, such as the measurement-device-independent (MDI) QCC protocols. MDI QCC, which is inspired by MDI QKD \cite{Pirandola2012,mdi_qkd}, removes all detector side channels, and is therefore secure and practical. The MDI QCC protocols proposed so far include polarization-encoding protocol \cite{pol_qcc}, phase-matching protocol \cite{pm_qcc}, $W$-state protocols \cite{w_qcc_1,w_qcc_2}, and asynchronous protocols \cite{lu2024repeaterlike,xie2024multifield}. 

For the polarization-encoding protocol \cite{pol_qcc}, the low probability of obtaining a successful GHZ-state projection measurement event due to multifold coincidence detection will induce large statistical fluctuation with finite data size in the case of high channel loss, thus reducing the key rate and communication distance. To solve this problem, in this work, we theoretically improve the previous protocol with four-intensity decoy-state method \cite{4_intensity_qkd,4_intensity_qkd_2}, which allows us to reliably estimate the parameters and thus significantly improve the key rate with finite data size compared to the original theoretical proposal \cite{pol_qcc}, and experimentally present a three-user quantum communication network using the polarization-encoding MDI QCC protocol \cite{pol_qcc} as shown in Fig. 1, in which users randomly send polarization-encoding optical pulses to the detection node for GHZ-state measurement and establish multipartite quantum correlation by successful projection measurement events. 

The MDI QCC protocols have a promising prospect of performing multiuser and large-scale quantum communication. However, the experimental research on the MDI QCC protocols is still lacking so far, which makes it difficult to establish practical MDI QCC network without preliminary fundamental work. Our work takes an important step toward the realization of the practical MDI QCC network.

\section{Protocol}
The four-intensity decoy-state protocol has been widely used in two-party MDI QKD experiments due to its ability to significantly improve the communication distance and key rate. We apply the four-intensity decoy-state method to the original MDI QCC protocol and perform theoretical calculations. The simulation results show a significant improvement in key rate and communication distance compared to the three-intensity protocol under the same conditions. In our four-intensity MDI QCC protocol, three users randomly and independently prepare optical pulses with an average photon number of $\mu_z$ as signal states with probability $p_z$ and with an average photon number of $\mu_x$, $\mu_y$, and $\mu_o=0$ as decoy states with probability $p_x$, $p_y$, $p_o$, respectively, where the signal states are encoded on the polarization $Z$ basis and the nonvacuum decoy states are encoded on the polarization $X$ basis. The encoded optical pulses are sent to the GHZ-state analyzer \cite{ghz_analyzer_2,ghz_analyzer_1} for GHZ-state measurement, where the states $|\Phi^+\rangle=(|HHH\rangle+|VVV\rangle)/\sqrt{2}$ and $|\Phi^-\rangle=(|HHH\rangle-|VVV\rangle)/\sqrt{2}$ are identified. The conference key rate is given by \cite{pol_qcc,pm_qcc}:

\begin{eqnarray}
    R&=&p_z^3\left\{ \mu_z^3 e^{-3\mu_z} Y_{111}^Z \left[1-H(e_{111}^{PZ}) \right]\right.\nonumber\\ 
    &\;& -Q_Z f \textrm{max}\left[H(e^Z_{\textrm{AB}}), H(e^Z_{\textrm{AC}}), H(e^Z_{\textrm{BC}}) \right] \Big\}, 
\end{eqnarray}
where $H(x)=-x\log_2{x}-(1-x)\log_2(1-x)$ is the binary Shannon entropy function; $Y_{111}^Z$ and $e_{111}^{PZ}$ are the yield and phase error rate of the single-photon components of $Z$-basis pulses, respectively; $Q_Z$ is the overall gain of $Z$-basis pulses; $e^Z_{\textrm{AB}}$, $e^Z_{\textrm{AC}}$, and $e^Z_{\textrm{BC}}$ are the quantum bit error rate of $Z$-basis pulses for Alice-Bob, Alice-Charlie, and Bob-Charlie; $f=1.16$ is the error correction efficiency.

\begin{figure}
\includegraphics[width=\linewidth]{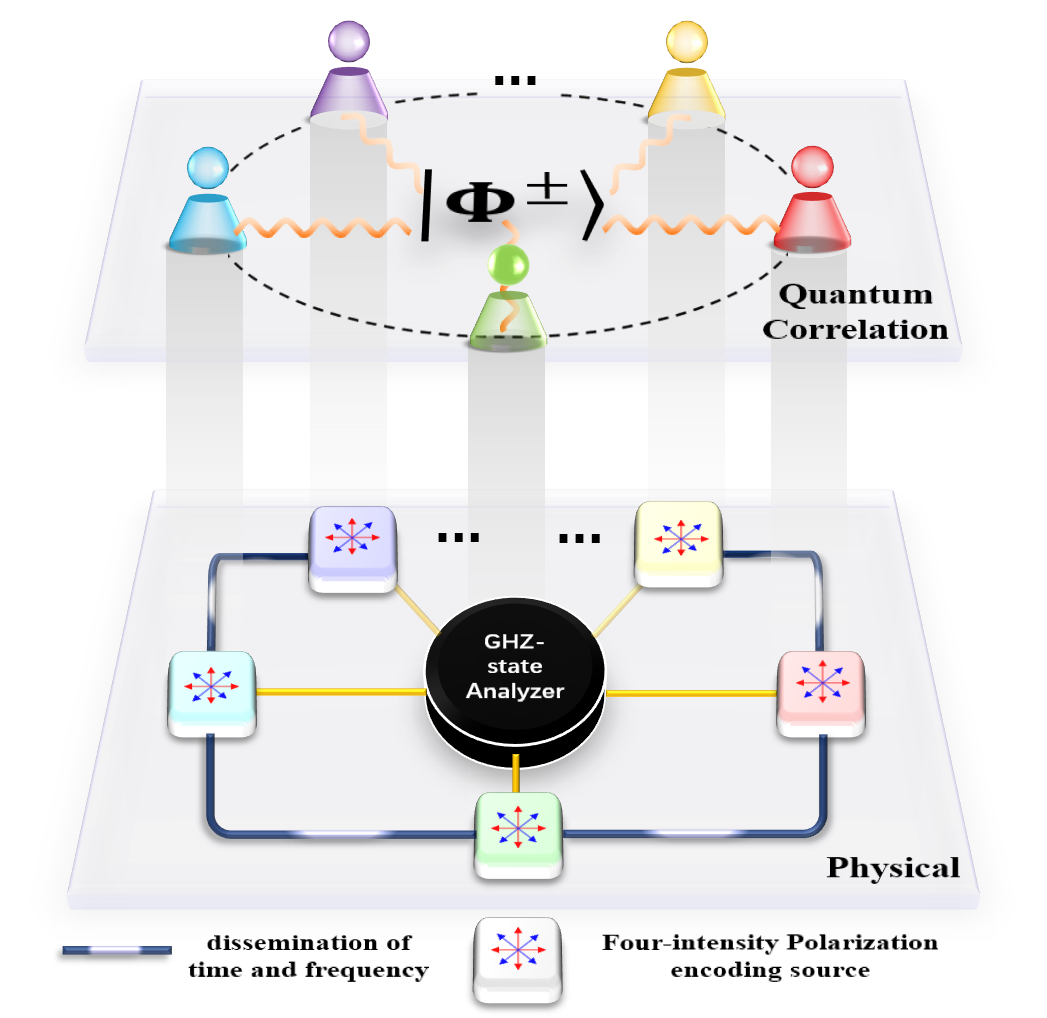}
\caption{Scheme of polarization encoding measurement-device-independent quantum cryptographic conferencing. In physical layer, multiple users participating in conference randomly prepare polarization-encoded four-intensity pulses with the help of dissemination of time and frequency for synchronization and frequency difference suppression. They send pulses to the GHZ-state analyzer at untrusted detection node for GHZ-state projection measurements. In quantum correlation layer, GHZ-state correlation among all users is generated from successful GHZ-state projection measurement events.}
\end{figure}

\section{Experiment}
\begin{figure*}
\includegraphics[width=\linewidth]{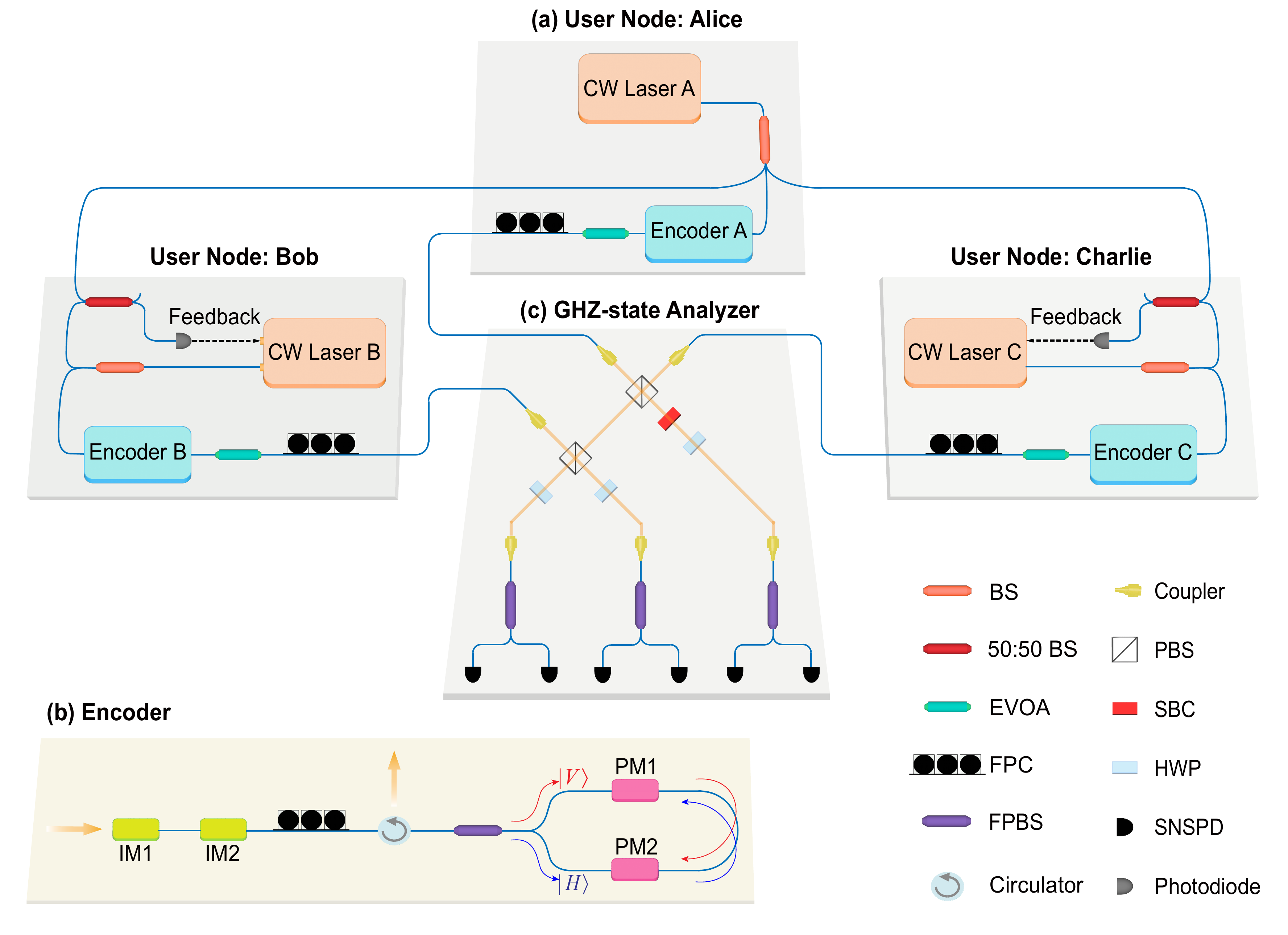}
\caption{\label{fig:device}Experimental setup of MDI QCC. (a) The encoding system to generate polarization qubits for three users, A, B, and C. Each user uses a continuous-wave (cw) laser as the light source. The frequency difference between Alice-Bob and Alice-Charlie is suppressed by means of the feedback control with the beat signal. Decoy states and polarization qubits are realized using an encoder. (b) Detailed structure of the encoder. The four different polarization states are generated by modulating the relative phase between $H$ and $V$ component of incident pulses in a Sagnac loop. (c) GHZ-state analyzer for polarization qubits; see text for details. Abbreviation of the components: BS, beam splitter; EVOA, electrical variable optical attenuator; FPC, fiber polarization controller; IM, intensity modulator; PM, phase modulator; PBS, polarization beam splitter; HWP, half-wave plate; SBC, Soleil-Babinet compensator; SNSPD, superconducting nanowire single-photon detector.}
\end{figure*}

The experimental setup of the MDI QCC is shown in Fig. 2, which consists of the polarization-encoding optical pulse transmitters held by three users and the GHZ-state analyzer at the untrusted detection node. In the user node, three legitimate users Alice, Bob, and Charlie use three independent narrow linewidth cw lasers (A, B, and C) as light sources, whose frequency can be tuned by external voltage controls. User B and user C each split 5\% of their light to interfere with that from user A and perform frequency stabilization feedback control. The remaining light of all three users is used for encoding. In the frequency feedback control system, Alice interferes her light with Bob’s and Charlie’s on a $50:50$ beam splitter for frequency beating, respectively. According to the beat signal detected by the photodiode, we feedback onto Bob’s and Charlie’s lasers to regulate the frequency difference between laser A and laser B, as well as that between laser A and laser C within 5 MHz for more than 30 hours. 

Another important part of the user node is the encoder. Each encoder can be divided into two parts. In the intensity-encoding part, the first intensity modulator (IM1) chops the cw light into a series of pulses with a repetition frequency of 250 MHz and a duration of about 200 ps. The second intensity modulator (IM2) implements the intensity modulation for the four-intensity decoy-state protocol. The optical pulses then enter the polarization encoder with the initial polarization state $(|H\rangle+|V\rangle)/\sqrt{2}$ \cite{pol_mod_1,pol_mod_2}. The polarization encoder generates different polarization states by modulating the relative phase between the $H$ and $V$ components of the incident light by the two phase modulators (PMs). By adjusting the relative delay of the electrical signals applied to the PMs, we modulate only the phase of the $V$ component passing clockwise through the loop and keep the phase of $H$ component unchanged. We set the amplitude of the electrical signal on PM1 to $V_{\pi}$ and on the PM2 to $V_{\pi/2}$ so that four relative phases ($0,\pi/2,\pi,3\pi/2$) can be generated to implement polarization encoding. The fidelity of all four polarization states can reach more than 0.99.

After encoding, the electrical variable optical attenuators (EVOAs) are used to attenuate the optical pulses to single-photon level and simulate channel loss. Then the optical pulses are sent into the GHZ-state analyzer. We rotate the polarization state to $Z$ and $X$ basis and compensate the polarization change during propagation by the fiber polarization controllers (FPCs). In our GHZ-state analyzer, we use fiber optical elements including three fiber polarization beam splitters (PBSs) for the part after the half-wave plates (HWPs). For the rest of the analyzer, we use bulk optical elements such as PBSs and HWPs. The angle of the birefringence optical axis of the HWPs on each optical path is set to 22.5$^{\circ}$. Especially considering the additional phase shift introduced by the PBSs, we add a Soleil-Babinet compensator (SBC) on one of the optical paths in the GHZ-state analyzer for phase compensation \cite{qcc_eb_exp_2}.

\begin{figure*}
\includegraphics[width=\linewidth]{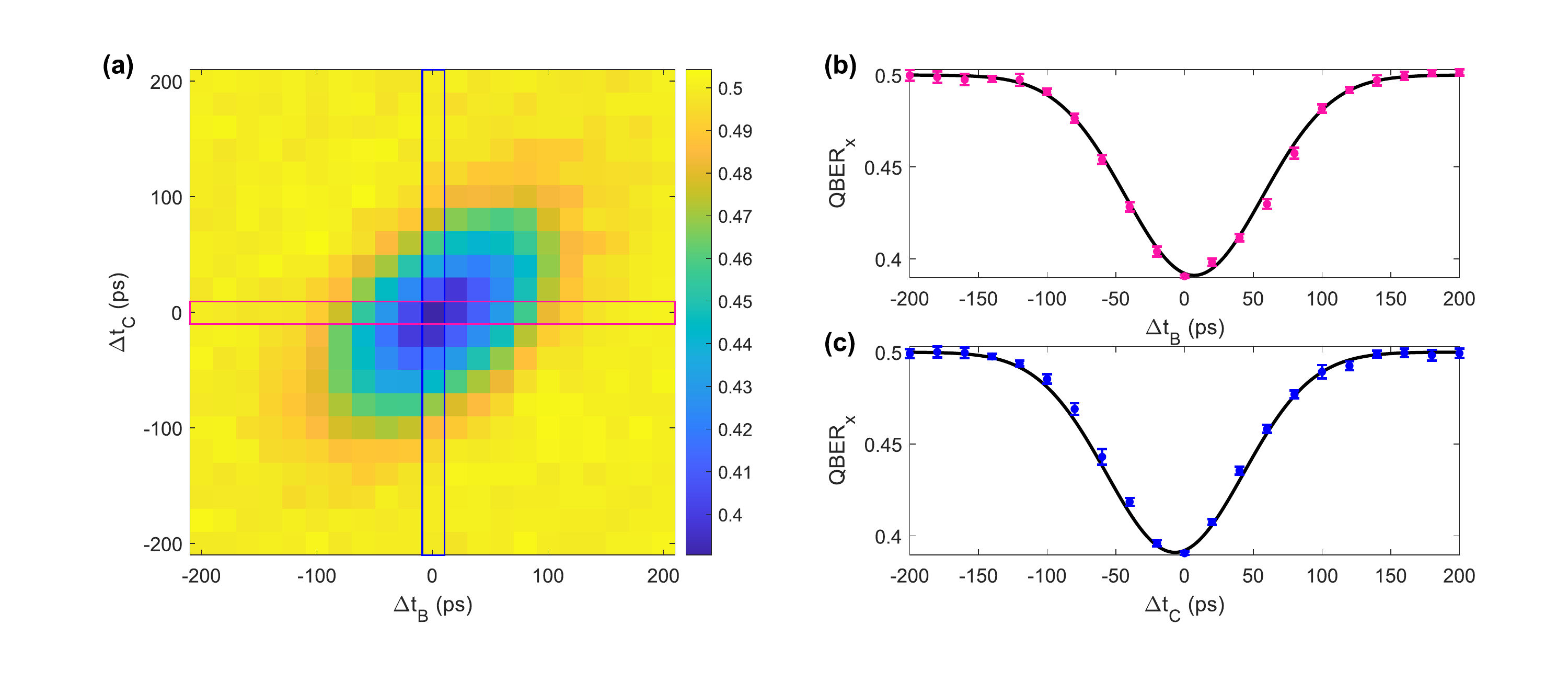}
\caption{\label{qber}Quantum bit error rate (QBER) of GHZ-state measurement. (a) $QBER_X$ when scanning the relative delay time of Bob and Charlie jointly. (b), (c) $QBER_X$ when scanning the relative delay time of Bob and Charlie in the case of $\Delta t_{C}=0$ [red slice in (a)] and $\Delta t_{B}=0$ [blue slice in (a)], respectively. The black curve is the fitting curve.}
\end{figure*}

To implement the MDI QCC protocol, we first perform the GHZ-state measurement experiment \cite{ghz_exp_1,ghz_exp_2,ghz_exp_3}. When scanning the relative delay time between two users, the threefold coincidence counts corresponding to the probability of being projected to $|\Phi^{\pm}\rangle$ states show the dip or peak, and we refer to this generalized Hong-Ou-Mandel (HOM) interference \cite{hom_1} phenomenon as GHZ-HOM interference. The ideal visibility of GHZ-HOM interference with weak coherent state is 25\%, resulting in ideal 37.5\% $QBER_X$ \footnote{See Supplemental Material at http://link.aps.org/supplemental/10.1103/PhysRevLett.134.040802 for the detailed analysis of the GHZ-HOM interference (Sec. I), which includes Ref. \cite{hom_2}}. We perform the GHZ-HOM experiment with independent weak coherent states, where three users randomly send $X$-basis polarization states with the same average photon number. Then we fix the delay time of Alice’s pulses and scan the delay time of Bob’s and Charlie’s pulses. The measurement results are shown in Fig. 3 and we get $QBER_X=39.10\pm0.09\%$, according to which the average visibility is about $21.80\pm0.18\%$.

Then in the process of conference key generation, we perform four-intensity MDI QCC protocol in three different attenuation scenarios. The optimized intensity and probability parameters selected in the experiments are $\mu_z=0.100$, $\mu_x=0.0281$, $\mu_y=0.152$, and the corresponding probabilities $p_z=0.33$, $p_x=0.51$, $p_y=0.09$. The average
detection efficiency of the superconducting nanowire single-photon detectors (SNSPDs) is about 80\% and the average of dark count rate is about 250 Hz. We perform the experiments with an overall attenuation of three users of about 14.1, 17.8, and 21.5 dB, which includes the total channel loss of three users, the detection efficiency of SNSPDs, and the efficiency of coincidence time window. With an accumulation time of $8\times 10^4$ s and a total number of pulses of about $1.99\times10^{13}$, we obtain the conference key rates of about 7.54, 1.17 and 0.097 bits per second (bps), respectively after decoy-state analysis based on the Chernoff bound method \cite{4_intensity_qkd_2,chernoff_bound,PhysRevA.95.012333}, in which we take the failure probability to be $\epsilon=10^{-10}$. The data points and theoretical key rate curves are shown in Fig. 4. We compare the key rate of the three-intensity protocol under the same conditions. The results show that the four-intensity protocol significantly improves the conference key rate and communication distance with finite data size.

\begin{figure}
\includegraphics[width=\linewidth]{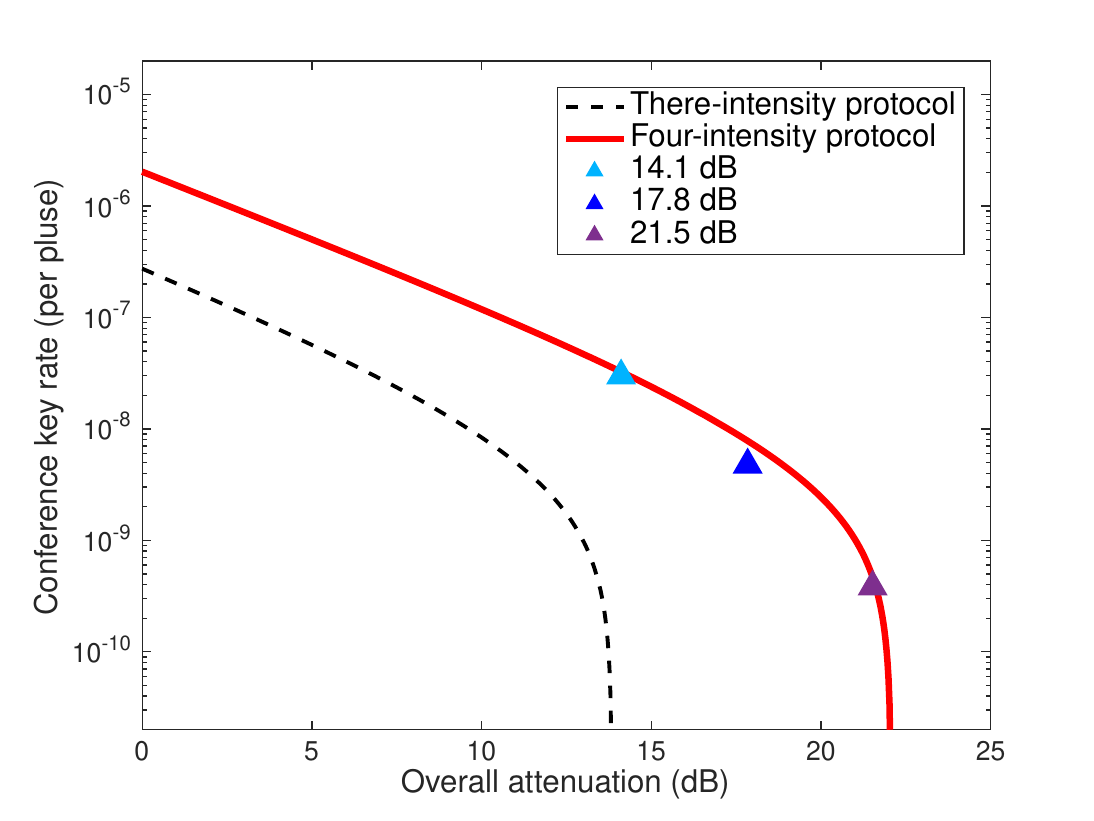}
\caption{\label{result}Results of conference key generation. In theoretical calculation, we set the misalignment error $e_d=2.25\%$, and take the interference visibility into consideration. We optimize the key rate of the three-intensity protocol under the same condition at the overall attenuation of 13.5 dB. The result shows that the $R_{\text{4-int}}/R_{\text{3-int}}\simeq 110$ at the overall attenuation of 13.5 dB. }
\end{figure}

\section{Discussion}
In summary, we experimentally demonstrate the MDI-QCC by the GHZ-state projection measurement. Using four-intensity decoy-state MDI-QCC protocol, our experimental work has significantly enhanced the secure conference key rate and communication distance, comparing to the original theoretical proposal. Our work is an important step toward the practical realization of multipartite quantum communication, represented by the transition from two-node, to multi-node experiments within the future quantum internet, Despite these important results, it is noted that the scale of the key rate of our MDI-QCC protocol for N users is $O(\eta^N)$. With current technology, although it is possible to realize 100-km scale MDI-QCC, covering most applications in metropolitan areas, this scaling may bring limitations to its scalability beyond 100 km. Currently, some single-photon MDI-QCC protocols such as W-state protocols \cite{w_qcc_1,w_qcc_2} and asynchronous protocols \cite{lu2024repeaterlike,xie2024multifield} have been proposed, which may overcome this scaling limitations. Based on our work, exploring experimental implementations of single-photon protocols is the next key task in the further research to establish practical MDI-QCC networks. 

~\

\textit{Note Added.}\textemdash We note that related experimental work has been reported in Ref. \cite{PhysRevLett.133.210803}.

\section{Acknowledgment}
This research was supported by the National Key Research and Development Program of China (Grant No. 2022YFE0137000), the Natural Science Foundation of Jiangsu Province (Grants No. BK20240006 and No. BK20233001), the Leading-Edge Technology Program of Jiangsu Natural Science Foundation (Grant No. BK20192001), the Innovation Program for Quantum Science and Technology (Grants No. 2021ZD0300700 and No. 2021ZD0301500), and the Fundamental Research Funds for the Central Universities (Grant No. 2024300324), and the Nanjing University-China Mobile Communications Group Co., Ltd. Joint Institute.

    
    \onecolumngrid
    \setcounter{section}{1} 
    \setcounter{figure}{0} 
    \setcounter{equation}{0}
	\renewcommand*{\thefigure}{S\arabic{figure}}
    \renewcommand*{\thetable}{S\arabic{table}}
    \renewcommand*{\thesection}{S\arabic{section}}
    \renewcommand*{\theequation}{S\arabic{equation}}
	\counterwithout{equation}{section}
    \numberwithin{equation}{section}

\newpage
\section*{Supplemental Materials}
\section{S1: Hong-Ou-Mandel interference in GHZ-state measurement}

In the MDI-QKD experiments, HOM interference \cite{hom_1} is often used to verify the spectral and temporal indistinguishability of photons from two users. Similarly, in the MDI-QCC experiment, the probability of being projected to GHZ-states shows a peak or dip while scanning the relative delay time. In this section, we theoretically analyze the GHZ-HOM interference with the weak coherent state based on the method given in \cite{hom_2}.

\subsection{S1.1: Theory of GHZ-HOM interference with weak coherent states}

Assume that the coherent state has the Gaussian temporal and spectral distribution centered at $t_0$ and $\omega_0$: 
 
\begin{equation}
    \begin{split}
    \left|\Psi\left(t_0\right)\right\rangle_{\mathrm{C}}&=\left(\frac{2 \Gamma}{\pi}\right)^{\frac{1}{4}} \int_{-\infty}^{\infty}\left|\sqrt{\alpha_{t-t_0}} e^{i \theta_{t-t_0}}\right\rangle_C \mathrm{d} t=\left(\frac{2 \Gamma}{\pi}\right)^{\frac{1}{4}} \int_{-\infty}^{\infty}\left|\sqrt{\alpha e^{-2\Gamma (t-t_0)^2}} e^{i\theta^0+i\omega_0 (t-t_0) }\right\rangle_C \mathrm{d} t,	
    \end{split}
\end{equation}		
in which $\alpha$ is the average photon number; $\theta^0$ is the random phase; $\Gamma$ is the linewidth parameter; $\omega_0$ and $t_0$ are the central frequency and time respectively. 

\begin{figure}[H]
\centering
\includegraphics[width=0.5\textwidth]{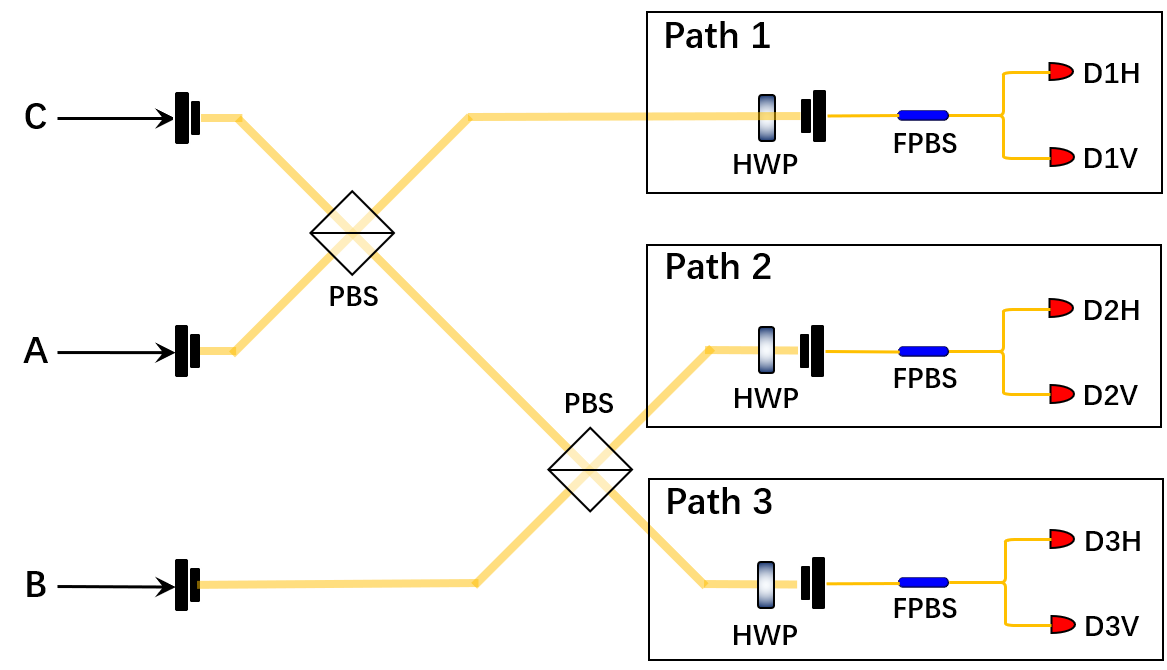}
\caption{\label{ghz-state-analyzer}GHZ-state analyzer}
\end{figure}

Without loss of generality, we assume that three users prepare the polarization state $|+\rangle$ with the average photon number of $\mu,\nu$ and $\kappa$, respectively. i.e. the input states are $\left|\sqrt{\mu}e^{i\varphi_A}\right\rangle_+\left|\sqrt{\nu}e^{i\varphi_B}\right\rangle_+\left|\sqrt{\kappa}e^{i\varphi_C}\right\rangle_+$. 
Taking into account the temporal distribution, the quantum states arriving at the six detectors after passing through the GHZ-state analyzer are as shown in Figure \ref{ghz-state-analyzer}:

\begin{equation}
    \left|\psi\right\rangle_{D 1 H / D 1 V}=\left({\frac{2 \Gamma}{\pi}}\right)^{1 / 4} \int_{-\infty}^{+\infty}\left|e^{i \varphi_A}\left(\frac{\sqrt{\mu_{t-t_A}}}{2} \pm \frac{\sqrt{\nu_{t-t_B}}}{2} e^{i \varphi_{A B}}\right)\right\rangle \mathrm{d} t,
\end{equation}

\begin{equation}
    \left|\psi\right\rangle_{D 2 H / D 2 V}=\left({\frac{2 \Gamma}{\pi}}\right)^{1 / 4} \int_{-\infty}^{+\infty}\left|e^{i \varphi_B}\left(\frac{\sqrt{\nu_{t-t_B}}}{2} \pm \frac{\sqrt{\kappa_{t-t_C}}}{2} e^{i \varphi_{B C}}\right)\right\rangle \mathrm{d} t,
\end{equation}

\begin{equation}
    \left|\psi\right\rangle_{D 3 H / D 3 V}=\left({\frac{2 \Gamma}{\pi}}\right)^{1 / 4} \int_{-\infty}^{+\infty}\left|e^{i \varphi_C}\left(\frac{\sqrt{\kappa_{t-t_C}}}{2} \pm \frac{\sqrt{\mu_{t-t_A}}}{2} e^{i \varphi_{C A}}\right)\right\rangle \mathrm{d} t,
\end{equation}
where the global phase $\varphi_{A,B,C}=\varphi^0_{A,B,C}+\omega_{A,B,C}(t-t_{A,B,C})$ are random and $\varphi_{AB,BC,CA}$ are the relative phase between users AB, BC and CA. The phase randomized coherent state with the average photon number $\alpha$ is the mixed state of the photon number state:

\begin{equation}
    \rho_{\text{PR}}=\frac{1}{2 \pi} \int_0^{2 \pi}\left|\sqrt{\alpha} e^{i \phi}\right\rangle\left\langle\sqrt{\alpha} e^{i \phi}\right|\mathrm{d} \phi=\sum_{n=0}^{+\infty} \frac{e^{-\alpha} \alpha^n}{n !}\left| n\right\rangle\langle n|.
\end{equation}
Given that the dark count rate is $p_d$ and the detection efficiency is $\eta$, the click probability of a detector is:

\begin{equation}
    P=1-\mathrm{e}^{-\alpha}\left(1-p_d\right) \sum_{n=0}^{+\infty} \frac{\alpha^n}{n !}(1-\eta)^n=1-\left(1-p_d\right) e^{-\alpha \eta}.
\end{equation}
Assuming that $p_d=0$ and $\eta=1$, if $\alpha\ll 1$, the click probability above can be approximated as: 

\begin{equation}\label{eq.1.7}
    P\simeq 1-(1-\alpha)=\alpha.
\end{equation}
It shows that the click probability of a phase-randomized weak coherent state is approximately equal to its average photon number.  
According to the eq. (\ref{eq.1.7}), for example, the click probability of a single detector D1H or D1V can be expanded as:
	
\begin{equation}
    \begin{split}
        \mathrm{P}_{D 1 H/D1V}=\sqrt{\frac{\Gamma}{8 \pi}} &\int_{-\infty}^{+\infty}\left[\mu e^{-2 \Gamma\left(t-t_A\right)^2}+\nu e^{-2 \Gamma\left(t-t_B\right)^2}\pm 2 \sqrt{\mu \nu} e^{-2 \Gamma\left(t-\frac{t_A+t_B}{2}\right)^2-\frac{\Gamma\Delta t^2_{AB}}{2}} \cos \left(\Delta\omega_{AB}t+\phi_{AB}\right)\right] \mathrm{d} t,
    \end{split}
\end{equation}
where $\phi_{AB}=\Delta\varphi^0_{AB}+\omega_At_A-\omega_Bt_B$ is the random phase. In the same way we can obtain the click probabilities of the remaining four detectors. After calculating the integration, we have:
\begin{equation}
    \mathrm{P}_{D 1 H/D1V}=\frac{1}{4}(\mu+\nu)\pm\frac{1}{2} \sqrt{\mu \nu} e^{-\frac{\Gamma\Delta t^2_{AB}}{2}-\frac{\Delta\omega_{AB}^2}{8\Gamma}} \cos \left(\Delta\phi_{A B}\right),
\end{equation}
\begin{equation}
    \mathrm{P}_{D 2 H/D2V}=\frac{1}{4}(\kappa+\nu)\pm\frac{1}{2} \sqrt{\kappa \nu} e^{-\frac{\Gamma\Delta t^2_{BC}}{2}-\frac{\Delta\omega_{BC}^2}{8\Gamma}} \cos \left(\Delta\phi_{BC}\right),
\end{equation}
\begin{equation}
    \mathrm{P}_{D 3 H/D3V}=\frac{1}{4}(\mu+\kappa)\pm\frac{1}{2} \sqrt{\mu \kappa} e^{-\frac{\Gamma\Delta t^2_{CA}}{2}-\frac{\Delta\omega_{CA}^2}{8\Gamma}} \cos \left(\Delta\phi_{CA}\right).
\end{equation}
We assume that $\mu=\nu=\kappa$ and the frequency difference between two users is sufficiently small, i.e. $\Delta\omega^2/8\Gamma \simeq 0 $. Then the 3-fold coincidence count rate is the product of the click probabilities of the corresponding detectors, i.e. the probabilities of being projected to the $|\Phi^+\rangle$ state and the $|\Phi^-\rangle$ state are: 
\begin{equation}\label{eq.1.12}
    P_{|\Phi^+\rangle}(\phi_{AB},\phi_{CA})=P_{D1H}P_{D2H}P_{D3H}+P_{D1H}P_{D2V}P_{D3V}+P_{D1V}P_{D2H}P_{D3V}+P_{D1V}P_{D2V}P_{D3H},
\end{equation}
\begin{equation}\label{eq.1.13}
    P_{|\Phi^-\rangle}(\phi_{AB},\phi_{CA})=P_{D1V}P_{D2V}P_{D3V}+P_{D1V}P_{D2H}P_{D3H}+P_{D1H}P_{D2V}P_{D3H}+P_{D1H}P_{D2H}P_{D3V}.
\end{equation}
Considering that the random phase is uniformly distributed on $[0,2\pi]$, the final result can be obtained by taking the mean of eq. (\ref{eq.1.12}) and eq. (\ref{eq.1.13}): 

\begin{equation}
    \begin{split}
        P_{|\Phi^+\rangle}&=\frac{1}{4\pi^2}\int_{0}^{2\pi}\mathrm{d}\phi_{AB}\int_{0}^{2\pi}\mathrm{d}\phi_{CA}{P_{|\Phi^+\rangle}(\phi_{AB},\phi_{CA})}=\frac{\mu^3}{2}\left[1+{\color{black}\frac{1}{4}}e^{-\frac{\Gamma}{2}\left(\Delta t_{AB}^2+\Delta t_{AC}^2+\Delta t_{BC}^2 \right)} \right],
    \end{split}  
\end{equation}

\begin{equation}
    \begin{split}
        P_{|\Phi^-\rangle}&=\frac{1}{4\pi^2}\int_{0}^{2\pi}\mathrm{d}\phi_{AB}\int_{0}^{2\pi}\mathrm{d}\phi_{CA}{P_{|\Phi^-\rangle}(\phi_{AB},\phi_{CA})}=\frac{\mu^3}{2}\left[1-{\color{black}\frac{1}{4}}e^{-\frac{\Gamma}{2}\left(\Delta t_{AB}^2+\Delta t_{AC}^2+\Delta t_{BC}^2 \right)} \right].
    \end{split}  
\end{equation}
From the two equations above, we can conclude that the theoretical visibility for coincidence dip and peak of GHZ-HOM interference are as follows:
\begin{equation}
    V_{\text{dip}}=\left(\overline{P}-\underline{P} \right)/{\overline{P}}=0.25,
\end{equation}
\begin{equation}
    V_{\text{peak}}=\left(\overline{P}-\underline{P} \right)/{\underline{P}}=0.25,
\end{equation}
where $\overline{P}$ and $\underline{P}$ is the supremum and infimum of the projection probability. Generally, in the case of $V_{\text{dip}}=V_{\text{peak}}=V$, the $QBER_X$ is:
\begin{equation}
    QBER_X=\frac{P_{|\Phi^-\rangle}}{P_{|\Phi^+\rangle}+P_{|\Phi^-\rangle}}=\frac{1}{2}\left[1-Ve^{-\frac{\Gamma}{2}\left(\Delta t_{AB}^2+\Delta t_{AC}^2+\Delta t_{BC}^2 \right)}\right].
\end{equation}
When the light pulses coincide exactly in the time domain, i.e. $\Delta t_{AB}=\Delta t_{AC}=\Delta t_{BC}=0$, we have $QBER_X=(1-V)/2$, which means that $QBER_X=0.375$ when $V=0.25$.

\subsection{S1.2: Experiment of GHZ-HOM interference}
\subsubsection{Two-photon HOM interference}
In order to align the optical pulses of the three users in the time domain, we first perform the two-photon HOM interference experiment. The results are shown in the figure below:
\begin{figure}[H]
\centering
\includegraphics[width=\textwidth]{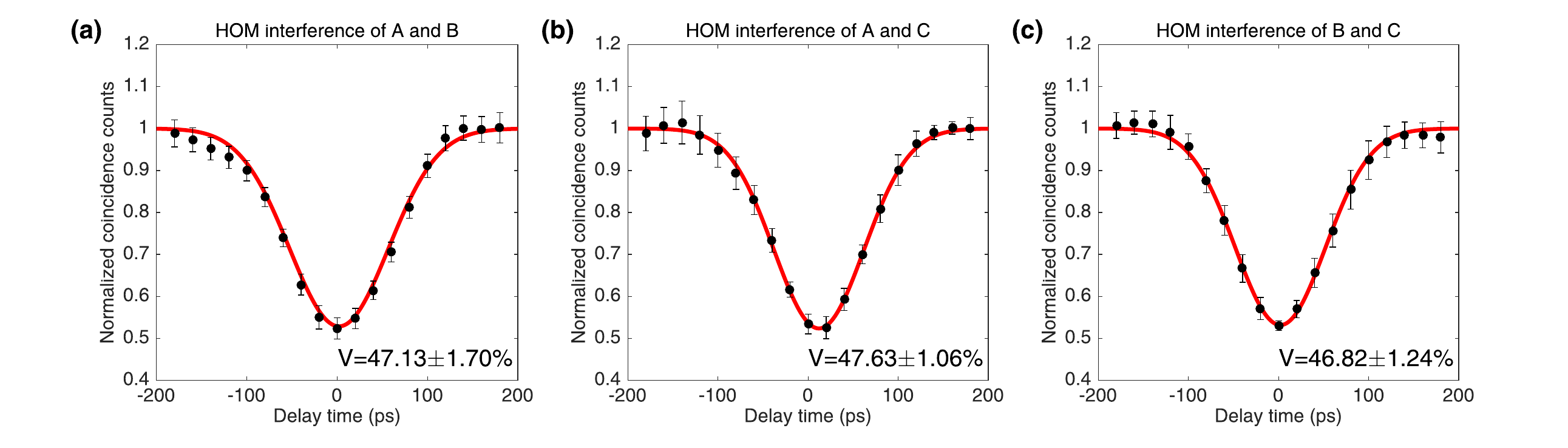}
\caption{\label{hom_2user}{Two-photon HOM interference.} (a) user A and user B; (b) user A and user C; (c) user B and user C. The black points are measurement data; the red lines are the fitting curves.}
\end{figure}

\subsubsection{Compensation of the phase shift}
We assume that the additional relative phase shift introduced by the polarization beam splitters on the $|V\rangle$ component for users A, B and C is $\gamma_1,\gamma_2$ and $\gamma_3$ respectively. The click probabilities of six detectors can be modified as follows:

\begin{equation}
    \mathrm{P}_{D 1 H/D1V}=\frac{1}{4}(\mu+\nu)\pm\frac{1}{2} \sqrt{\mu \nu} e^{-\frac{\Gamma\Delta t^2_{AB}}{2}-\frac{\Delta\omega_{AB}^2}{8\Gamma}} \cos \left(\Delta\phi_{A B}+\gamma_1\right),
\end{equation}
\begin{equation}
    \mathrm{P}_{D 2 H/D2V}=\frac{1}{4}(\kappa+\nu)\pm\frac{1}{2} \sqrt{\kappa \nu} e^{-\frac{\Gamma\Delta t^2_{BC}}{2}-\frac{\Delta\omega_{BC}^2}{8\Gamma}} \cos \left(\Delta\phi_{BC}+\gamma_2\right),
\end{equation}
\begin{equation}
    \mathrm{P}_{D 3 H/D3V}=\frac{1}{4}(\mu+\kappa)\pm\frac{1}{2} \sqrt{\mu \kappa} e^{-\frac{\Gamma\Delta t^2_{CA}}{2}-\frac{\Delta\omega_{CA}^2}{8\Gamma}} \cos \left(\Delta\phi_{CA}+\gamma_3\right),
\end{equation}

Repeating the same calculation as above, we can get the visibility:

\begin{equation}
    V=\frac{1}{4}\cos(\gamma_1+\gamma_2+\gamma_3).
\end{equation}
To eliminate the effect of additional phase on the interference, we add a Soleil-Babinet Compensator (SBC) in the optical path to adjust $\gamma_1$ to satisfy the condition $\gamma_1+\gamma_2+\gamma_3=2k\pi (k\in\mathbb{Z})$. In the experiment, three users randomly send two polarization states in X-basis with the equal probability. After aligning the pulses in the time and frequency domain, we scan the distance of the SBC from 0 to 20 mm in a 1mm steps. Each data point corresponds to a measurement lasting 50 s with coincidence time window of about 500 ps. The result is shown in Figure \ref{sbc_scan}:

\begin{figure}[H]
\centering
\includegraphics[width=0.4\textwidth]{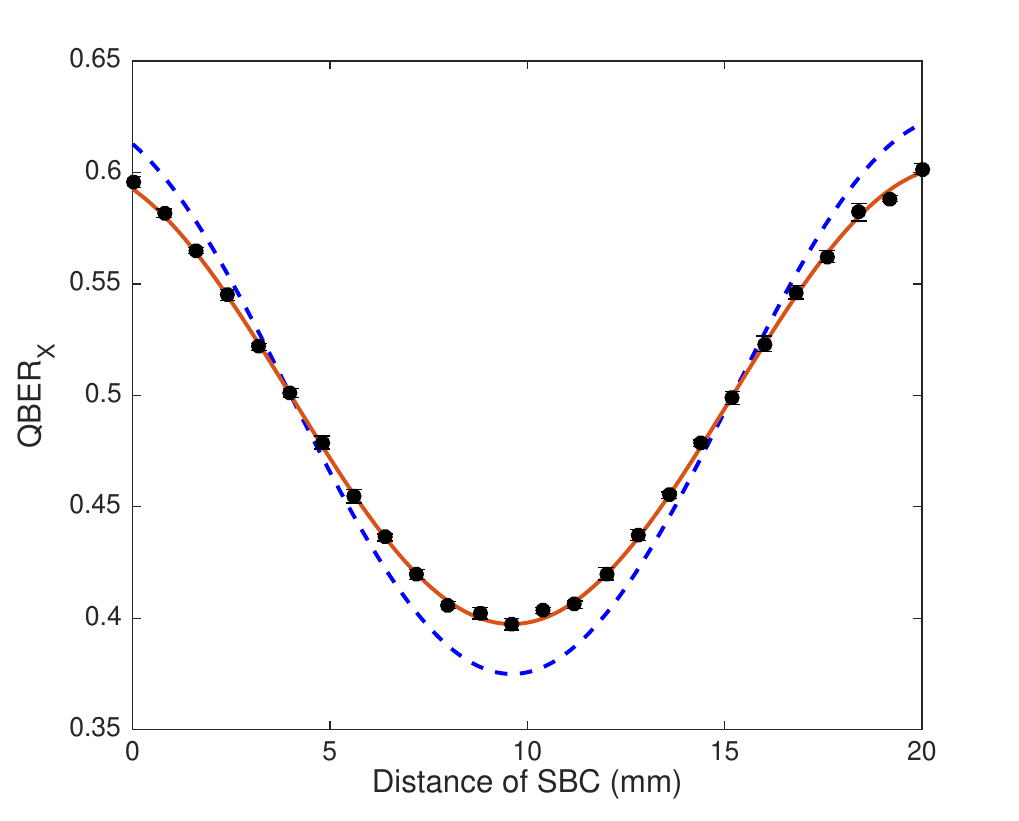}
\caption{\label{sbc_scan}{${QBER}_{{X}}$ versus the distance of SBC.} The black points are measurement data; the red line is the fitting curve; the blue dashed line is the theoretical curve. Since the compensation phase of the SBC is linearly related to the distance, $QBER_X$ is cosine-related to $\gamma_1$, which is consistent with the theoretical result.}
\end{figure}

\subsubsection{Results of GHZ-HOM interference}
\begin{figure}[H]
\centering
\includegraphics[width=\textwidth]{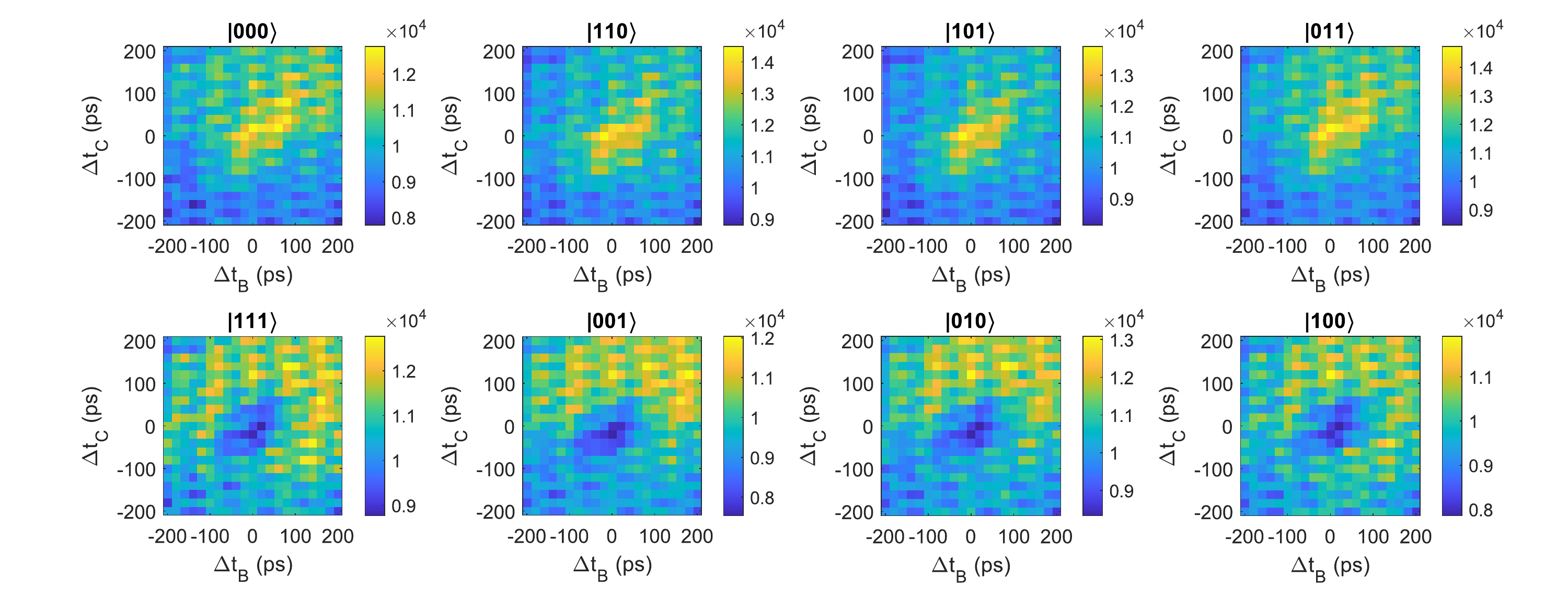}
\caption{\label{original1}{Original coincidence counts projected to $|\Phi^+\rangle$ of eight states in X-basis}}
\end{figure}

\begin{figure}[H]
\centering
\includegraphics[width=\textwidth]{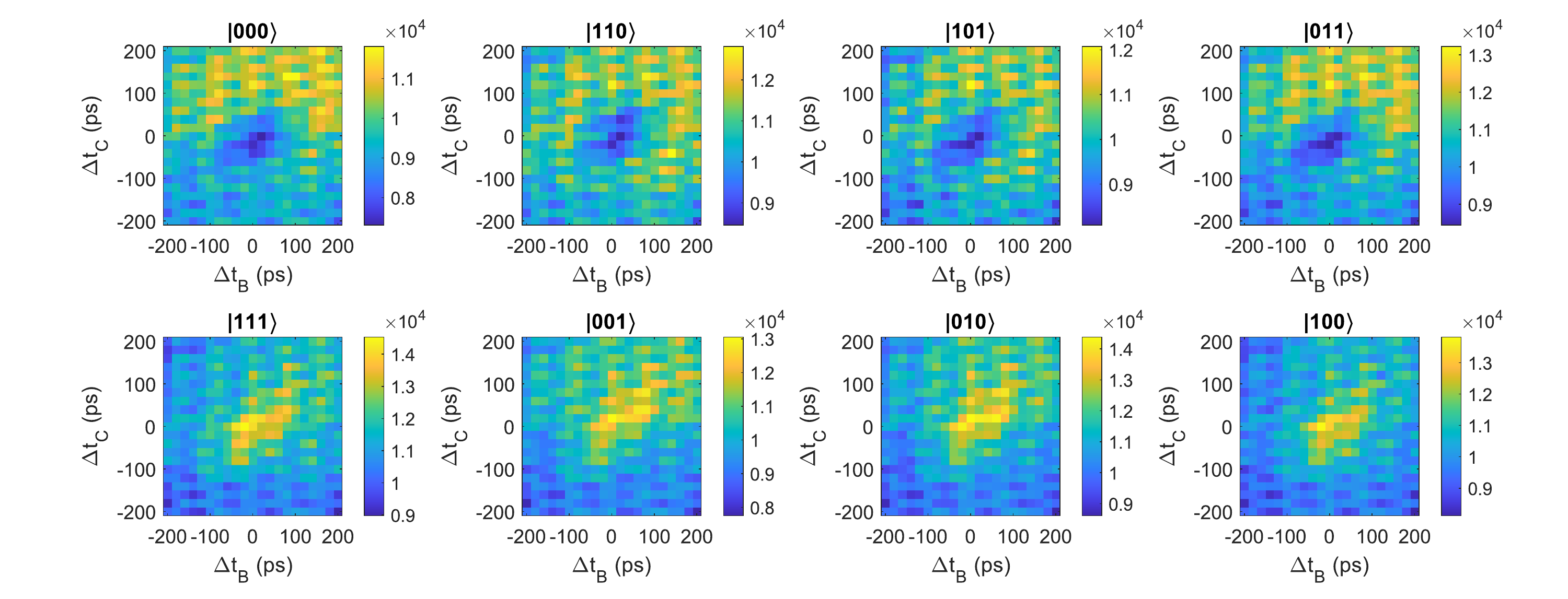}
\caption{\label{original2}{Original coincidence counts projected to $|\Phi^-\rangle$ of eight states in X-basis}}
\end{figure}

After adjusting the SBC to compensate phase shift, we scan the relative delay time between user AB and AC from -200 ps to 200 ps in the 20 ps steps. Each data point corresponds to a measurement lasting 60 seconds with coincidence time window of about 500 ps. The original 3-fold coincidence counts of the GHZ-HOM interference with different input states are shown in Figure \ref{original1} and Figure \ref{original2}.  

Considering the overall fluctuation while scanning the delay time, we calculate the proportional coincidence counts as:
\begin{equation}
\Bar{N}^{|\Phi^{+/-}\rangle}_{|\psi\rangle}=\frac{1}{C_{\text{fit}}}\frac{16N^{|\Phi^{+/-}\rangle}_{|\psi\rangle}}{N_{\text{total}}},
\end{equation}
where $N^{|\Phi^{+/-}\rangle}_{|\psi\rangle}$ is the coincidence counts projected onto $|\Phi^{+}\rangle$ or $|\Phi^{-}\rangle$ state with the input state $|\psi\rangle$; $N_{\text{total}}$ is the total coincidence counts of all input states; and the $C_{\text{fit}}$ is the normalization constant obtained by the fitting results. 

\begin{figure}[H]
\centering
\includegraphics[width=\textwidth]{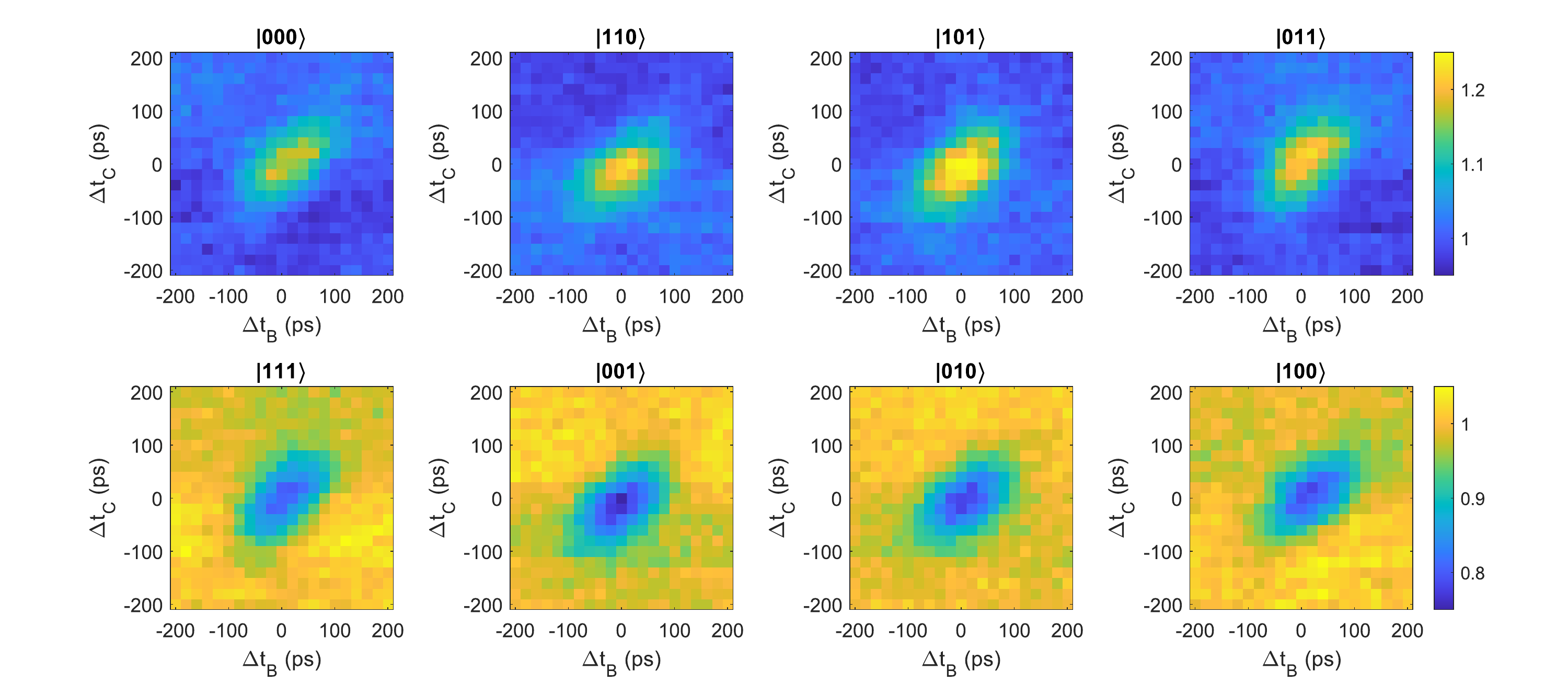}
\caption{\label{process1}{Processed coincidence counts projected to $|\Phi^+\rangle$ of eight input states in X-basis}}
\end{figure}

\begin{figure}[H]
\centering
\includegraphics[width=\textwidth]{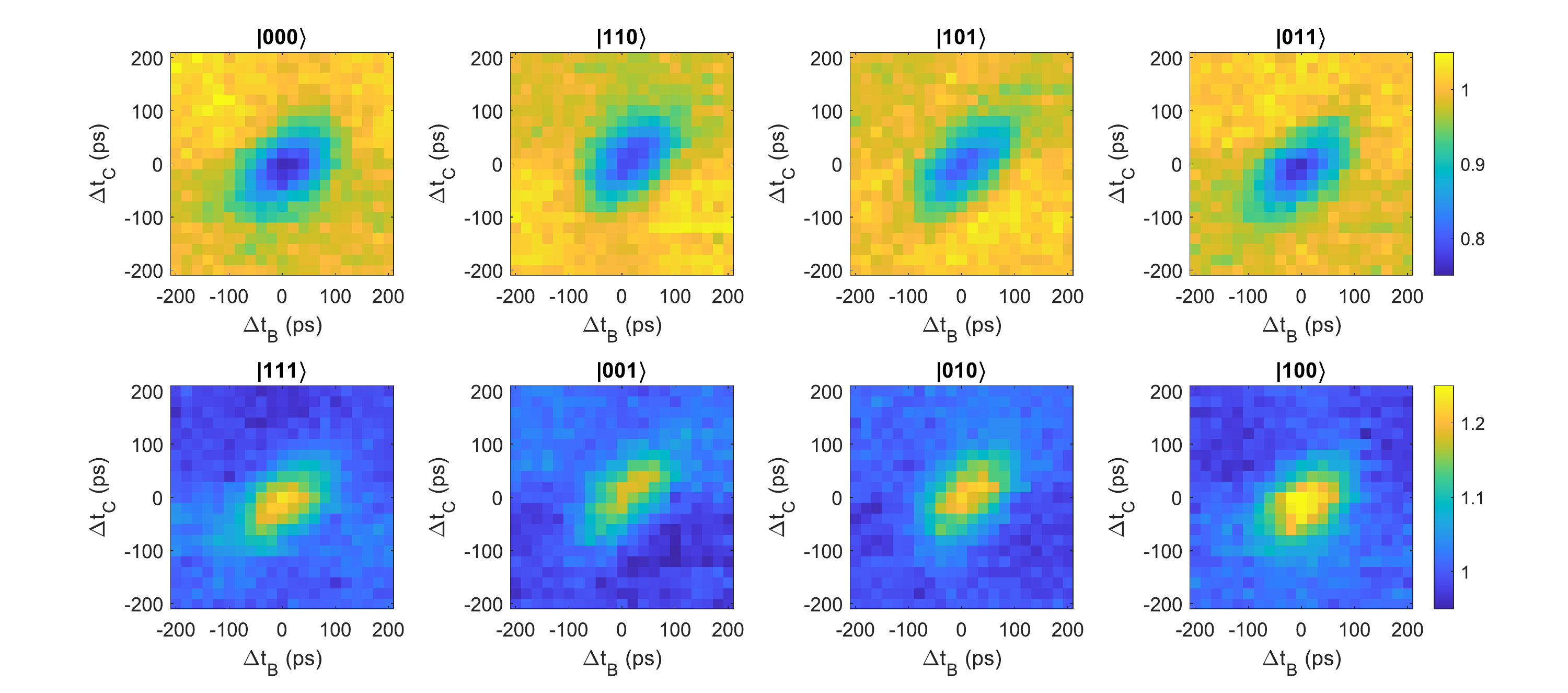}
\caption{\label{process2}{Processed coincidence counts projected to $|\Phi^-\rangle$ of eight input states in X-basis}}
\end{figure}

\begin{figure}[H]
\centering
\includegraphics[width=\textwidth]{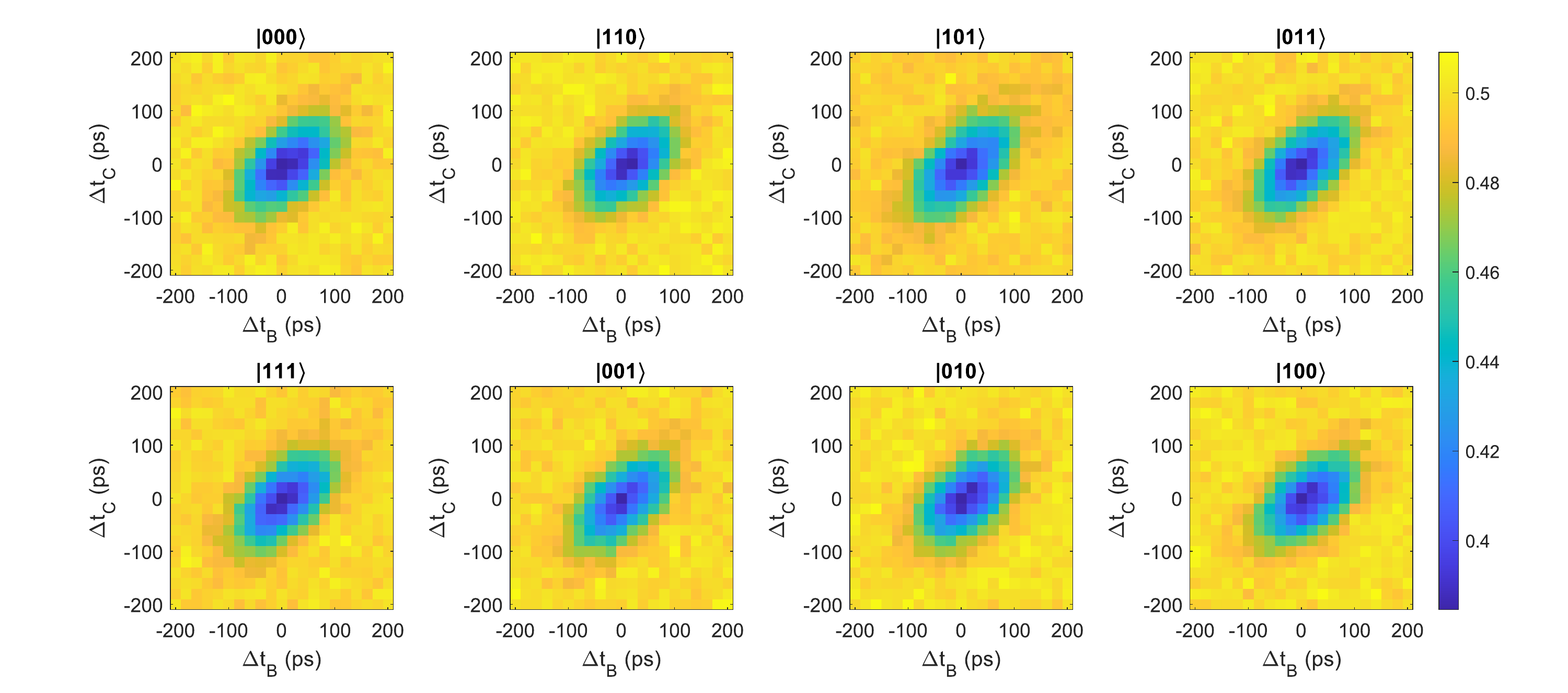}
\caption{\label{qber_all}{$QBER_X$ of eight input states in X-basis}}
\end{figure}

The processed coincidence counts and $QBER_X$ as a function of $\Delta t_{B}$ and $\Delta t_{C}$ are shown in Figure \ref{process1}, Figure \ref{process2} and Figure \ref{qber_all}. The visibilities are obtained by fitting the $QBER_X$ data based on the theoretical equation eq. (1.18) and the results are shown in Table \ref{hom_table}.

\begin{table}[H]
\centering
\begin{tabular}{@{}ccc@{}}
\toprule
Input state   & Visibility ($\%$)         & $QBER_X$ ($\%$)           \\ \midrule
$|000\rangle$ & $21.72\pm 0.33$ & $39.14\pm 0.17$ \\
$|001\rangle$ & $21.20\pm 0.35$ & $39.40\pm 0.18$ \\
$|010\rangle$ & $22.26\pm 0.30$ & $38.87\pm 0.15$ \\
$|011\rangle$ & $21.87\pm 0.33$ & $39.07\pm 0.17$ \\
$|100\rangle$ & $22.83\pm 0.35$ & $38.59\pm 0.18$ \\
$|101\rangle$ & $22.24\pm 0.36$ & $38.88\pm 0.18$ \\
$|110\rangle$ & $21.47\pm 0.29$ & $39.27\pm 0.15$ \\
$|111\rangle$ & $21.09\pm 0.31$ & $39.46\pm 0.16$ \\ \bottomrule
\end{tabular}
\caption{\label{hom_table}{Visibility of GHZ-HOM interference and $QBER_X$ of eight input states}}
\end{table}

\section{S2: Frequency and polarization stability}
\begin{figure}[H]
    \centering
    \includegraphics[width=0.8\linewidth]{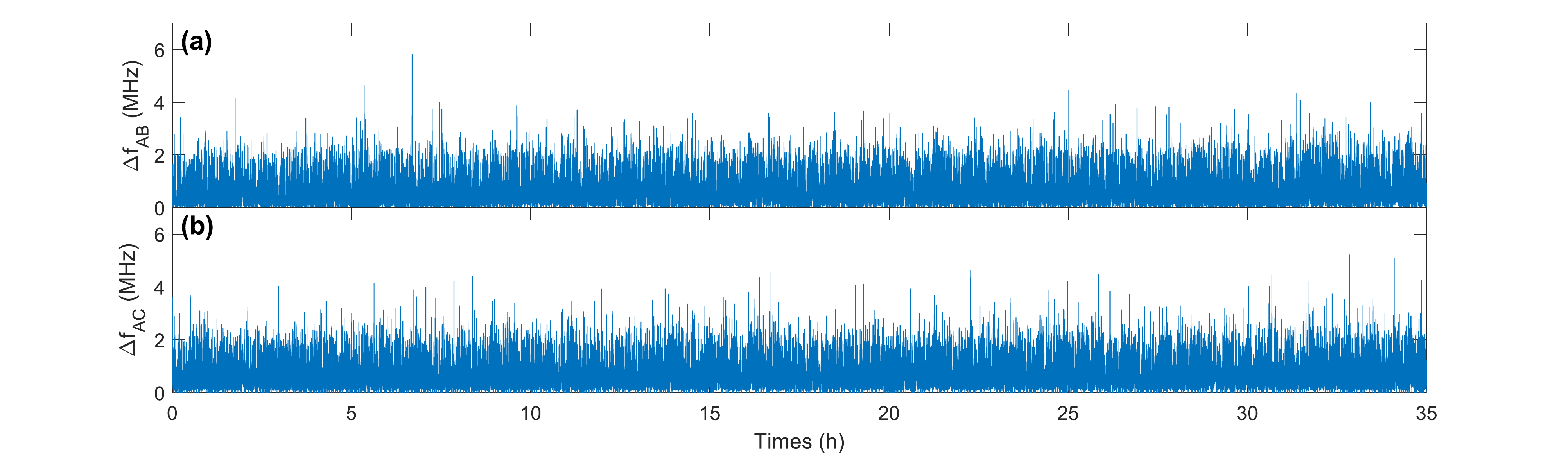}
    \caption{{Frequency difference between two users within $10^5$s. (a) $|\Delta f_{AB}|$, (b) $|\Delta f_{AC}|$ }.}
    \label{fig:freq_lock}
\end{figure}

\begin{figure}[H]
    \centering
    \includegraphics[width=0.8\linewidth]{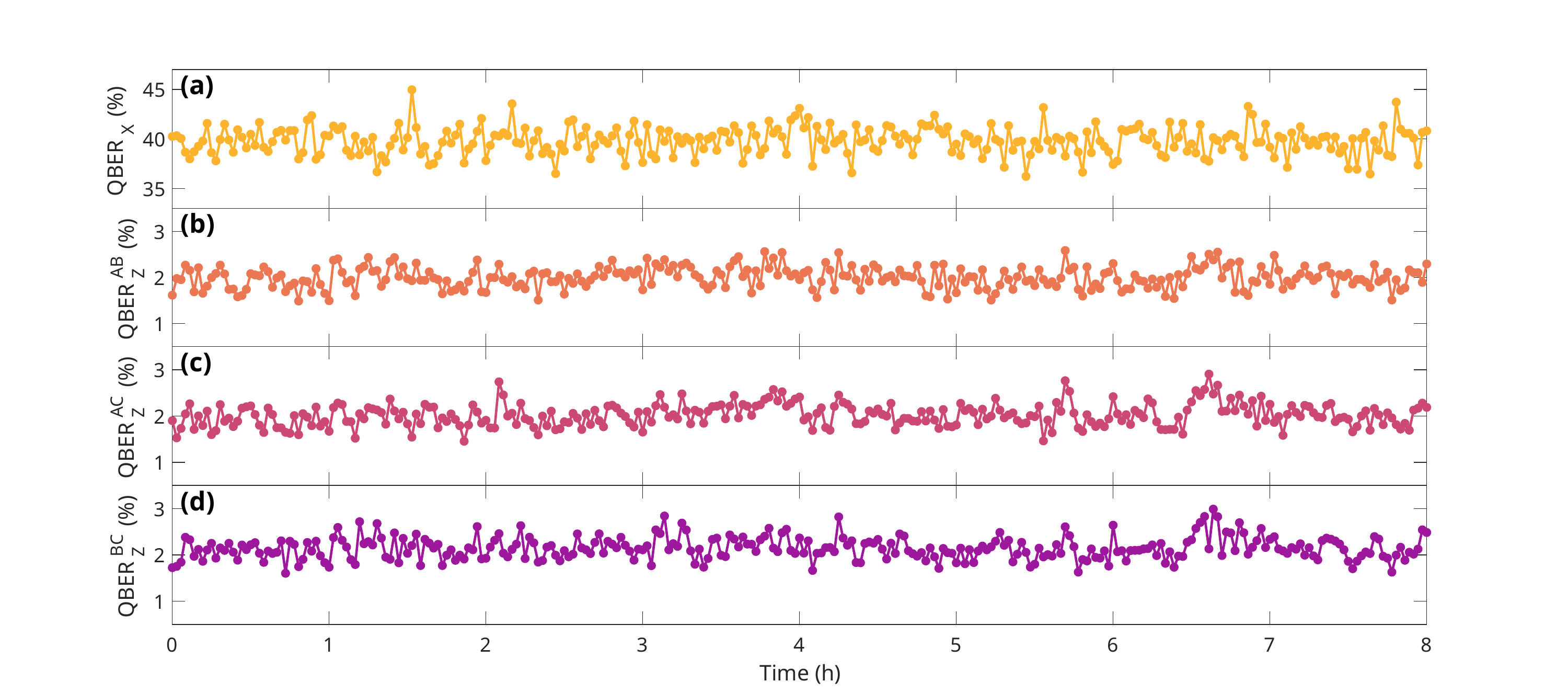}
    \caption{QBERs within 8 hours. For X-basis pulse pairs, (a) $QBER_X=39.78\pm1.40 \%$. For Z-basis pulse pairs, (b) $e_Z^{AB}=2.00\pm0.24\%$, (c) $e_Z^{AC}=2.02\pm0.24\%$, (d) $e_Z^{BC}=2.14\pm0.25\%$}
    \label{fig:qber_lt}
\end{figure}

The overall stability of the system is critical due to the long period of time required to accumulate data in the experiment. On the one hand, the frequency difference between the independent lasers can be suppressed to less than 5 MHz by feedback control and can be kept stable within $10^5$ seconds as shown in Figure \ref{fig:freq_lock}. On the other hand, the $QBER_X$ and $QBER_Z$ of the system can be kept at a low level within about 8 hours, indicating that our polarization encoders and the GHZ-state analyzer also can be kept stable over a relatively long time as shown in Figure \ref{fig:qber_lt}. Besides, the degree of polarization (DOP) and fidelity of four states prepared by three users are shown in Table \ref{polmod}. In the experiment, we check the $QBER_X$ and $QBER_Z$ every 1000 seconds and adjust the system if the error rate is too high to generate the secure key. We adjusted the system on average $1\sim 3$ times during the measurement time of about 23 hours.
\begin{table}[H]
\centering
\begin{tabular}{|c|cc|cc|cc|cc|}
\hline
\multirow{3}{*}{} & \multicolumn{2}{c|}{$|D\rangle$}             & \multicolumn{2}{c|}{$|A\rangle$}                         & \multicolumn{2}{c|}{$|H\rangle$}                               & \multicolumn{2}{c|}{$|V\rangle$}                               \\ \cline{2-9} 
                  & \multicolumn{1}{c|}{$\phi_1=0$} & $\phi_2=0$ & \multicolumn{1}{c|}{$\phi_1=\pi$}     & $\phi_2=0$       & \multicolumn{1}{c|}{$\phi_1=0$}       & $\phi_2=\frac{\pi}{2}$ & \multicolumn{1}{c|}{$\phi_1=\pi$}     & $\phi_2=\frac{\pi}{2}$ \\ \cline{2-9} 
                  & \multicolumn{1}{c|}{$V_1=0$}        & $V_2=0$   & \multicolumn{1}{c|}{$V_1=V_{\pi}$}              & $V_2=0$         & \multicolumn{1}{c|}{$V_1=0$}              & $V_2=V_{\frac{\pi}{2}}$               & \multicolumn{1}{c|}{$V_1=V_{\pi}$}              & $V_2=V_{\frac{\pi}{2}}$
               \\ \cline{2-9} 
                  & \multicolumn{1}{c|}{DOP}        & Fidelity   & \multicolumn{1}{c|}{DOP}              & Fidelity         & \multicolumn{1}{c|}{DOP}              & Fidelity               & \multicolumn{1}{c|}{DOP}              & Fidelity       \\ \hline
user A            & \multicolumn{1}{c|}{$1.000(8)$}    & $1.000(3)$    & \multicolumn{1}{c|}{$0.987(3)$} & $0.996(1)$ & \multicolumn{1}{c|}{$0.980(7)$} & $0.995(2)$       & \multicolumn{1}{c|}{$0.991(2)$} & $0.997(1)$       \\ \hline
user B            & \multicolumn{1}{c|}{$1.000(3)$}    & $1.000(1)$    & \multicolumn{1}{c|}{$0.996(3)$} & $0.999(1)$ & \multicolumn{1}{c|}{$0.967(6)$} & $0.992(2)$       & \multicolumn{1}{c|}{$0.986(2)$} & $0.996(1)$       \\ \hline
user C            & \multicolumn{1}{c|}{$1.000(2)$}    & $1.000(1)$    & \multicolumn{1}{c|}{$0.995(3)$} & $0.998(1)$ & \multicolumn{1}{c|}{$0.969(7)$} & $0.992(2)$       & \multicolumn{1}{c|}{$0.995(2)$} & $0.998(1)$       \\ \hline
\end{tabular}
\caption{\label{polmod}{Degree of polarization (DOP) and fidelity of four polarization states output by the encoder}. We remark that in the experiment we set the polarization state $|D\rangle$ as reference}
\end{table}

\section{S3: Conference key rate of four-intensity MDI-QCC protocol}
    \setcounter{section}{3} 
    \setcounter{equation}{0}

The security of a QCC protocol is based on two conditions: correctness and secrecy, which is the same as QKD protocols \cite{qcc_review}. The security of the polarization-encoding MDI-QCC protocol with three-intensity decoy-state method has been proved in \cite{pol_qcc} based on the entanglement purification method. In this section, we derive the formulas for the decoy-state analysis of the four-intensity MDI-QCC protocol based on the method in \cite{4_intensity_qkd}. Then we analysis the security of the four-intensity decoy-state protocol and perform numerical simulations for the four-intensity protocol and the original three-intensity protocol. The results show that the four-intensity protocol significantly improves the key rate and the communication distance of the polarization-encoding MDI-QCC with finite data size.

\subsection{S3.1: Decoy state analysis}

In the four-intensity decoy-state protocol, each user uses four light sources with different intensities: the signal source $z$ with the average photon number of $\mu_z$ is encoded only on the Z-basis; the decoy sources $x$, $y$ with the average photon number of $\mu_x$, $\mu_y$ respectively are encoded on the X-basis, where $\mu_y>\mu_x$; the vacuum source with $\mu_o=0$. The density matrix of phase-randomized weak coherent state pulses generated by three non-vacuum light sources can be expressed as:

\begin{equation}
		\rho_z=\sum_{n=0}^{\infty}{u_n^z|n\rangle\langle n|}\quad \rho_x=\sum_{n=0}^{\infty}{u_n^x|n\rangle\langle n|}\quad 
		\rho_y=\sum_{n=0}^{\infty}{u_n^y|n\rangle\langle n|}\quad u=a,b,c.
	\end{equation}

In the calculation below we use $\langle\cdot\rangle$ to denote the expectation values. When three users use the $x$ source or $y$ source, the gains can be expressed as the sum of the yields of the components with different photon numbers:

\begin{equation}
		\langle Q_{xxx}\rangle =\sum_{i,j,k=0}{a^x_i b^x_j c^x_k\langle Y_{ijk}\rangle },
	\end{equation}
	\begin{equation}
	    \langle Q_{yyy}\rangle=\sum_{i,j,k=0}{a^y_i b^y_j c^y_k \langle Y_{ijk}\rangle}.
	\end{equation}
Then the gains including vacuum sources are:
\begin{equation}
    \begin{split}
        &\langle Q_{sso}\rangle =\sum_{i,j=0}{a^s_i b^s_j\langle Y_{ij0}\rangle },\quad
        \langle Q_{sos}\rangle =\sum_{i,k=0}{a^s_i c^s_k\langle Y_{i0k}\rangle },\quad
        \langle Q_{oss}\rangle =\sum_{j,k=0}{b^s_j c^s_k\langle Y_{0jk}\rangle },
    \end{split}
\end{equation}

\begin{equation}
    \begin{split}
        &\langle Q_{soo}\rangle =\sum_{i=0}{a^s_i\langle Y_{i00}\rangle },\quad
        \langle Q_{oso}\rangle =\sum_{j=0}{b_j^s\langle Y_{0j0}\rangle },\quad
        \langle Q_{oos}\rangle =\sum_{k=0}{c^s_k\langle Y_{00k}\rangle },
    \end{split}
\end{equation}

\begin{equation}
    \langle Q_{ooo}\rangle =\langle Y_{000}\rangle.
\end{equation}
in which $s=x,y$. Combining the above equations, we can get the gain excluding the part that includes the vacuum component: 
\begin{equation}
			\widetilde{Q}_{xxx}=\langle Q_{xxx}\rangle-\langle \mathcal{H}_{xxx}\rangle=\sum_{i,j,k=1}{a^x_i b^x_j c^x_k \langle Y_{ijk}\rangle}=a^x_1 b^x_1 c^x_1 \langle Y_{111}\rangle+a^x_1 b^x_2 c^x_1 \langle Y_{121}\rangle+\cdots,
	\end{equation}
	\begin{equation}
	    \widetilde{Q}_{yyy}=\langle Q_{yyy}\rangle-\langle \mathcal{H}_{yyy}\rangle=\sum_{i,j,k=1}{a^y_i b^y_j c^y_k \langle Y_{ijk}\rangle}=a^y_1 b^y_1 c^y_1 \langle Y_{111}\rangle+a^y_1 b^y_2 c^y_1 \langle Y_{121}\rangle+\cdots,
	\end{equation}	
where the excluded parts are:

\begin{equation}
	    \begin{split}
	        \langle \mathcal{H}_{xxx}\rangle= a^x_0\langle Q_{oxx}\rangle&+b^x_0 \langle Q_{xox}\rangle+c^x_0 \langle Q_{xxo}\rangle\\&-a^x_0 b^x_0 \langle Q_{oox}\rangle-a^x_0 c^x_0 \langle Q_{oxo}\rangle-b^x_0 c^x_0 \langle Q_{xoo}\rangle+a^x_0 b^x_0 c^x_0 \langle Q_{ooo}\rangle,
	    \end{split}
	\end{equation}
	
	\begin{equation}
	    \begin{split}
	        \langle \mathcal{H}_{yyy}\rangle= a^y_0\langle Q_{oyy}\rangle&+b^y_0 \langle Q_{yoy}\rangle+c^y_0 \langle Q_{yyo}\rangle\\&-a^y_0 b^y_0 \langle Q_{ooy}\rangle-a^y_0 c^y_0 \langle Q_{oyo}\rangle-b^y_0 c^y_0 \langle Q_{yoo}\rangle+a^y_0 b^y_0 c^y_0 \langle Q_{ooo}\rangle.
	    \end{split}
	\end{equation}
 
To estimate the yield of the single-photon component, we need to eliminate the terms $\langle Y_{211}\rangle$, $\langle Y_{121}\rangle$, and $\langle Y_{112}\rangle$ from the above two formulas. We assume that three users use the same light sources in the protocol, i.e. $a_n^s=b_n^s=c_n^s$. These terms can be eliminated as: 
\begin{equation}
        \begin{split}
            a^y_1 b^y_2 c^y_1 \widetilde{Q}_{xxx}-a^x_1 b^x_2 c^x_1 \widetilde{Q}_{yyy}&=\left[a_1^x a_1^y c_1^x c_1^y\left(b_1^x b_2^y-b_2^x b_1^y\right)\right]\langle Y_{111}\rangle\\&+\sum_{\substack{m+n+p\ge5\\m\ge1,n\ge1,p\ge1}}{\left(a_1^y b_2^y c_1^y a_m^x b_n^x c_p^x-a_1^x b_2^x c_1^x a_m^y b_n^y c_p^y\right)\langle Y_{mnp}\rangle}. 
        \end{split}
	\end{equation}
For the weak coherent state, we have:
\begin{equation}
    \frac{u_k^y}{u_k^x}=\frac{e^{-\mu_y}\mu_y^k/k!}{e^{-\mu_x}\mu_x^k/k!}=e^{\mu_x-\mu_y}\left( \frac{\mu_y}{\mu_x}\right)^k.
\end{equation}
Since $\mu_x<\mu_y$, the coefficient of the $\langle Y_{mnp}\rangle$ term in the above formula can be rewritten as:
\begin{equation}
    \left(a_1^y b_2^y c_1^y a_m^x b_n^x c_p^x-a_1^x b_2^x c_1^x a_m^y b_n^y c_p^y\right)=a_1^x b_2^x c_1^x a_m^x b_n^x c_p^x\left(\frac{a_1^y}{a_1^x}\frac{b_2^y}{b_2^x}\frac{c_1^y}{c_1^x}-\frac{a_m^y}{a_m^x}\frac{b_n^y}{b_n^x}\frac{c_p^y}{c_p^x} \right).
\end{equation}
According to the formulas above, we have:	

\begin{equation}
	    \left(\frac{a_1^y}{a_1^x}\frac{b_2^y}{b_2^x}\frac{c_1^y}{c_1^x}-\frac{a_m^y}{a_m^x}\frac{b_n^y}{b_n^x}\frac{c_p^y}{c_p^x} \right)=e^{3(\mu_x-\mu_y)}\left[\left(\frac{\mu_y}{\mu_x}\right)^4-\left(\frac{\mu_y}{\mu_x}\right)^{m+n+p} \right]\le e^{3(\mu_x-\mu_y)}\left(\frac{\mu_y}{\mu_x}\right)^4\left[1-\left(\frac{\mu_y}{\mu_x}\right) \right]<0.
	\end{equation}
Thus the coefficients of $\langle Y_{mnp}\rangle$ are negative. Then we can obtain the lower bound of the single-photon yield of X-basis as:
\begin{equation}
		\langle Y_{111}^X\rangle \geq \langle Y_{111}^X\rangle^{L}(\mathcal{H})=\frac{a^y_1 b^y_2 c^y_1 \widetilde{Q}_{xxx}-a^x_1 b^x_2 c^x_1 \widetilde{Q}_{yyy}}{a_1^x a_1^y c_1^x c_1^y\left(b_1^x b_2^y-b_2^x b_1^y\right)}=\frac{S^L_{+}-S^U_{-}-e^{-3\mu_y}\mu_y^4\mathcal{H}}{e^{-3 \mu_{x}-3 \mu_{y}}\left(\mu_{x}^{3} \mu_{y}^{4}-\mu_{x}^{4} \mu_{y}^{3}\right)},
	\end{equation}
where
\begin{equation}
	    \mathcal{H}=\mathcal{H}_{xxx},
	\end{equation}
	
	\begin{equation}
	\begin{split}
	    {S}_{+}=e^{-3 \mu_{y}} \mu_{y}^{4}\langle Q_{x x x}\rangle&+e^{-3 \mu_{x}} \mu_{x}^{4}\left[e^{-3 \mu_{y}} \langle Q_{ooo}\rangle+\right.\left. e^{-\mu_{y}}\left(\langle Q_{oyy}\rangle+\langle Q_{yoy}\rangle+\langle Q_{yyo}\rangle\right)\right],
	\end{split}
	\end{equation}

	\begin{equation}
		{S}_{-}=e^{-3 \mu_{x}} \mu_{x}^{4}\left[\langle Q_{yyy}\rangle+e^{-2 \mu_{y}}\left(\langle Q_{yoo}\rangle+\langle Q_{oyo}\rangle+\langle Q_{ooy}\rangle\right)\right].
	\end{equation}

On the other hand, we can also use the same method to estimate the single-photon phase error rate with the bit error rate of $x$-source pulses. Considering that the polarization state of the vacuum state can be randomly assigned to the $|+\rangle$ state or the $|-\rangle$ state with equal probability in the statistical process, the single-photon bit error rate of the X-basis pulses can be estimated as follows:
\begin{equation}
		\left\langle E_{xxx}Q_{xxx}\right\rangle-\frac{1}{2}\langle \mathcal{H}_{xxx}\rangle=\sum_{j, k,l\ge1} a_{j}^{x} b_{k}^{x} c_l^x\left\langle e_{jkl}Y_{j k l}\right\rangle \geqslant a_{1}^{x} b_{1}^{x}c^x_1\left\langle e_{111}Y_{111}\right\rangle.
	\end{equation}
The upper bound of the single-photon quantum bit error rate of the X-basis pulses is:
\begin{equation}
		    \langle e_{111}^{BX}\rangle\leq \langle e_{111}^{BX}\rangle^U(\mathcal{H})= \frac{\left\langle E_{xxx}Q_{xxx}\right\rangle^U-\langle \mathcal{H}\rangle/2}{\mu_x^3 e^{-3\mu_x}\langle Y_{111}\rangle^{L}}.
\end{equation}

Finally, considering that the expectation values satisfy $\langle Y_{111}^X\rangle=\langle Y_{111}^Z\rangle$ and $\langle e_{111}^{BX}\rangle=\langle e_{111}^{PZ}\rangle$, the key rate of the MDI-QCC protocol can be represented as the function of $\mathcal{H}$:
\begin{equation}
    R(\mathcal{H})=p_z^3\left\{\mu_z^3e^{-3\mu_z}Y_{111}^Z\left[1-H\left(e_{111}^{PZ}\right)\right]-Q_{zzz}f\text{max}\left[H\left(e_{Z}^{AB}\right),H\left(e_{Z}^{AC}\right),H\left(e_{Z}^{BC}\right)\right] \right\}.
\end{equation}

\subsection{S3.2: Finite key size analysis with Chernoff bound}

In the experiment, the number of pulses is finite, hence we need to take the statistical fluctuation into consideration. To calculate the secure key rate, we need to estimate the real value of the single-photon yield $Y_{111}^Z$ and the phase error rate $e_{111}^{PZ}$ of the Z-basis pulses, which cannot be measured directly in the experiment. In our calculation, according to the formulas in Section 3.1, we first estimate the expectation values $\langle Y_{111}^X\rangle^L$ and $\langle e_{111}^{BX}\rangle^U$ with the Chernoff bound method \cite{chernoff_bound,PhysRevA.95.012333,4_intensity_qkd_2} and the joint constraints. Since the expectation values satisfy $\langle Y_{111}^X\rangle=\langle Y_{111}^Z\rangle$ and $\langle e_{111}^{BX}\rangle=\langle e_{111}^{PZ}\rangle$, we can use the Chernoff bound again to obtain the real values.

\subsubsection{Chernoff bound}
\quad
Let $X_1,X_2,\dots,X_n$ be the measurement results of $n$ independent random events. If the measurement is successful, record it as 1, otherwise, record it as 0. Let $\chi=\sum_{i=1}^{n}{X_i}$ be the sum of the measurement results and $\langle\chi \rangle=\mathbb{E}(\chi)$ be the expected value of $\chi$.

\begin{itemize}
    
\item[(1)] 
\textbf{From $\chi$ to $\langle\chi\rangle$.}
Given the measurement result $\chi$ and the failure probability $\varepsilon$, the confidence interval of expected value satisfying $\text{Pr}\{\langle\chi\rangle >\mathbb{E}^{L}(\chi)\}=\text{Pr}\{\langle\chi\rangle <\mathbb{E}^{U}(\chi)\}\le\varepsilon/2$ can be evaluated as:
\begin{equation}
\langle\chi\rangle^L=\mathbb{E}^{L}(\chi)=\frac{\chi}{1+\delta^{L}}, \quad
\langle\chi\rangle^U=\mathbb{E}^{U}(\chi)=\frac{\chi}{1-\delta^{U}},
\end{equation}
where $\delta^L$ and $\delta^U$ can be obtained by solving the equations below:
\begin{equation}
{\left[\frac{e^{\delta^{L}}}{\left(1+\delta^{L}\right)^{1+\delta^{L}}}\right]^{\chi /\left(1+\delta^{L}\right)}=\frac{1}{2} \varepsilon}, \quad
{\left[\frac{e^{-\delta^{U}}}{\left(1-\delta^{U}\right)^{1-\delta^{U}}}\right]^{\chi /\left(1-\delta^{U}\right)}=\frac{1}{2} \varepsilon}.
\end{equation}
Let $b=-\ln{(\varepsilon/2)}$, if $\chi\ge 6b$, we use approximation \cite{PhysRevA.95.012333}:
\begin{equation}
\delta^L=\delta^U=\frac{3b+\sqrt{8b\chi+b^2}}{2(\chi-b)}.
\end{equation}
If the measurement result is 0, we have:
\begin{equation}
    \mathbb{E}^{L}(\chi)=0, \quad
    \mathbb{E}^{U}(\chi)=b.
\end{equation}

\item[(2)] 
\textbf{From $\langle\chi\rangle$ to $\chi$.}
Given the expected value $\langle\chi\rangle$ the confidence interval of the real value satisfying $\text{Pr}\{\chi\in [\chi^L,\chi^U] \}\ge1-\varepsilon$ can be evaluated as:
\begin{equation}
\chi^{L}=\mathbb{O}^L(\langle\chi\rangle)=(1-\delta'^U) \langle\chi\rangle, \quad
\chi^{U}=\mathbb{O}^U(\langle\chi\rangle)=(1+\delta'^L) \langle\chi\rangle,
\end{equation}
where $\delta^L$ and $\delta^U$ can be samely obtained by solving the equations below:
\begin{equation}
{\left[\frac{e^{\delta'^{L}}}{\left(1+\delta'^{L}\right)^{1+\delta'^{L}}}\right]^{\langle\chi\rangle}=\frac{1}{2} \varepsilon}, \quad
{\left[\frac{e^{-\delta'^{U}}}{\left(1-\delta'^{U}\right)^{1-\delta'^{U}}}\right]^{\langle\chi\rangle}=\frac{1}{2} \varepsilon}.
\end{equation}
According to the results in \cite{PhysRevA.95.012333}, in the calculation we use the formula:
\begin{equation}
    \delta'^L=\delta'^U =\frac{b+\sqrt{b^{2}+8b\langle\chi\rangle}}{2 \langle\chi\rangle}.
\end{equation}
where $b=-\ln{(\varepsilon/2)}$. 

\end{itemize}

\subsubsection{Joint constraints}
\quad
In computing $S_+$ and $S_-$, similar to the method in \cite{4_intensity_qkd,4_intensity_qkd_2}, we introduce a set of joint constraints and perform linear programming to find the minimum and maximum values, respectively. In the inequalities below we denote $n$ as measurement values and $N$ as the number of pulses. 26 inequalities are introduced to calculate the minimum value of $S_+$.  

The joint constraints for all 5 variables are: 
\begin{equation}
    N_{xxx}\langle Q_{xxx}\rangle+N_{oyy}\langle Q_{oyy}\rangle+N_{yoy}\langle Q_{yoy}\rangle+N_{yyo}\langle Q_{yyo}\rangle+N_{ooo}\langle Q_{ooo}\rangle\ge
        \mathbb{E}^{L}(n_{xxx}+n_{oyy}+n_{yoy}+n_{yyo}+n_{ooo}).
\end{equation}

The joint constraints for 4 variables are: 
    \begin{gather}
       N_{oyy}\langle Q_{oyy}\rangle+N_{yoy}\langle Q_{yoy}\rangle+N_{yyo}\langle Q_{yyo}\rangle+N_{ooo}\langle Q_{ooo}\rangle\ge
        \mathbb{E}^{L}(n_{oyy}+n_{yoy}+n_{yyo}+n_{ooo}),\\
        N_{xxx}\langle Q_{xxx}\rangle+N_{yoy}\langle Q_{yoy}\rangle+N_{yyo}\langle Q_{yyo}\rangle+N_{ooo}\langle Q_{ooo}\rangle\ge
        \mathbb{E}^{L}(n_{xxx}+n_{yoy}+n_{yyo}+n_{ooo}),\\
        N_{xxx}\langle Q_{xxx}\rangle+N_{oyy}\langle Q_{oyy}\rangle+N_{yyo}\langle Q_{yyo}\rangle+N_{ooo}\langle Q_{ooo}\rangle\ge
        \mathbb{E}^{L}(n_{xxx}+n_{oyy}+n_{yyo}+n_{ooo}),\\
        N_{xxx}\langle Q_{xxx}\rangle+N_{oyy}\langle Q_{oyy}\rangle+N_{yoy}\langle Q_{yoy}\rangle+N_{ooo}\langle Q_{ooo}\rangle\ge
        \mathbb{E}^{L}(n_{xxx}+n_{oyy}+n_{yoy}+n_{ooo}),\\
        N_{xxx}\langle Q_{xxx}\rangle+N_{oyy}\langle Q_{oyy}\rangle+N_{yoy}\langle Q_{yoy}\rangle+N_{yyo}\langle Q_{yyo}\rangle\ge
        \mathbb{E}^{L}(n_{xxx}+n_{oyy}+n_{yoy}+n_{yyo}).
    \end{gather}

The joint constraints for 3 variables are:     
    \begin{gather}
        N_{yoy}\langle Q_{yoy}\rangle+N_{yyo}\langle Q_{yyo}\rangle+N_{ooo}\langle Q_{ooo}\rangle\ge
        \mathbb{E}^{L}(n_{yoy}+n_{yyo}+n_{ooo}),\\
        N_{oyy}\langle Q_{oyy}\rangle+N_{yyo}\langle Q_{yyo}\rangle+N_{ooo}\langle Q_{ooo}\rangle\ge
        \mathbb{E}^{L}(n_{oyy}+n_{yyo}+n_{ooo}),\\
        N_{oyy}\langle Q_{oyy}\rangle+N_{yoy}\langle Q_{yoy}\rangle+N_{ooo}\langle Q_{ooo}\rangle\ge
        \mathbb{E}^{L}(n_{oyy}+n_{yoy}+n_{ooo}),\\
        N_{oyy}\langle Q_{oyy}\rangle+N_{yoy}\langle Q_{yoy}\rangle+N_{yyo}\langle Q_{yyo}\rangle\ge
        \mathbb{E}^{L}(n_{oyy}+n_{yoy}+n_{yyo}),\\
        N_{xxx}\langle Q_{xxx}\rangle+N_{yyo}\langle Q_{yyo}\rangle+N_{ooo}\langle Q_{ooo}\rangle\ge
        \mathbb{E}^{L}(n_{xxx}+n_{yyo}+n_{ooo}),\\
        N_{xxx}\langle Q_{xxx}\rangle+N_{yoy}\langle Q_{yoy}\rangle+N_{ooo}\langle Q_{ooo}\rangle\ge
        \mathbb{E}^{L}(n_{xxx}+n_{yoy}+n_{ooo}),\\
        N_{xxx}\langle Q_{xxx}\rangle+N_{yoy}\langle Q_{yoy}\rangle+N_{yyo}\langle Q_{yyo}\rangle\ge
        \mathbb{E}^{L}(n_{xxx}+n_{yoy}+n_{yyo}),\\
        N_{xxx}\langle Q_{xxx}\rangle+N_{oyy}\langle Q_{oyy}\rangle+N_{ooo}\langle Q_{ooo}\rangle\ge
        \mathbb{E}^{L}(n_{xxx}+n_{oyy}+n_{ooo}),\\
        N_{xxx}\langle Q_{xxx}\rangle+N_{oyy}\langle Q_{oyy}\rangle+N_{yyo}\langle Q_{yyo}\rangle\ge
        \mathbb{E}^{L}(n_{xxx}+n_{oyy}+n_{yyo}),\\
        N_{xxx}\langle Q_{xxx}\rangle+N_{oyy}\langle Q_{oyy}\rangle+N_{yoy}\langle Q_{yoy}\rangle\ge
        \mathbb{E}^{L}(n_{xxx}+n_{oyy}+n_{yoy}).
    \end{gather}

The joint constraints for 2 variables are: 
	\begin{gather}
		N_{xxx}\langle Q_{xxx}\rangle+N_{oyy}\langle Q_{oyy}\rangle\ge
		\mathbb{E}^{L}(n_{xxx}+n_{oyy}),\\
		N_{xxx}\langle Q_{xxx}\rangle+N_{yoy}\langle Q_{yoy}\rangle\ge
		\mathbb{E}^{L}(n_{xxx}+n_{yoy}),\\
		N_{xxx}\langle Q_{xxx}\rangle+N_{yyo}\langle Q_{yyo}\rangle\ge
		\mathbb{E}^{L}(n_{xxx}+n_{yyo}),\\
		N_{xxx}\langle Q_{xxx}\rangle+N_{ooo}\langle Q_{ooo}\rangle\ge
		\mathbb{E}^{L}(n_{xxx}+n_{ooo}),\\
		N_{oyy}\langle Q_{oyy}\rangle+N_{yoy}\langle Q_{yoy}\rangle\ge
		\mathbb{E}^{L}(n_{oyy}+n_{yoy}),\\
		N_{oyy}\langle Q_{oyy}\rangle+N_{yyo}\langle Q_{yyo}\rangle\ge
		\mathbb{E}^{L}(n_{oyy}+n_{yyo}),\\
		N_{oyy}\langle Q_{oyy}\rangle+N_{ooo}\langle Q_{ooo}\rangle\ge
		\mathbb{E}^{L}(n_{oyy}+n_{ooo}),\\
		N_{yoy}\langle Q_{yoy}\rangle+N_{yyo}\langle Q_{yyo}\rangle\ge
		\mathbb{E}^{L}(n_{yoy}+n_{yyo}),\\
		N_{yoy}\langle Q_{yoy}\rangle+N_{ooo}\langle Q_{ooo}\rangle\ge
		\mathbb{E}^{L}(n_{yoy}+n_{ooo}),\\
		N_{yyo}\langle Q_{yyo}\rangle+N_{ooo}\langle Q_{ooo}\rangle\ge
		\mathbb{E}^{L}(n_{yyo}+n_{ooo}).
	\end{gather}

\quad
Similarly, for $S_-$, there are 11 inequalities. The joint constraints for all 4 variables are:
\begin{equation}
	N_{yyy}\langle Q_{yyy}\rangle+N_{yoo}\langle Q_{yoo}\rangle+N_{oyo}\langle Q_{oyo}\rangle+N_{ooy}\langle Q_{ooy}\rangle\le
	\mathbb{E}^{U}(n_{yyy}+n_{yoo}+n_{oyo}+n_{ooy}).
\end{equation}

The joint constraints for 3 variables are:  
\begin{gather}
	N_{yoo}\langle Q_{yoo}\rangle+N_{oyo}\langle Q_{oyo}\rangle+N_{ooy}\langle Q_{ooy}\rangle\le
	\mathbb{E}^{U}(n_{yoo}+n_{oyo}+n_{ooy}),\\
	N_{yyy}\langle Q_{yyy}\rangle+N_{oyo}\langle Q_{oyo}\rangle+N_{ooy}\langle Q_{ooy}\rangle\le
	\mathbb{E}^{U}(n_{yyy}+n_{oyo}+n_{ooy}),\\
	N_{yyy}\langle Q_{yyy}\rangle+N_{yoo}\langle Q_{yoo}\rangle+N_{ooy}\langle Q_{ooy}\rangle\le
	\mathbb{E}^{U}(n_{yyy}+n_{yoo}+n_{ooy}),\\
	N_{yyy}\langle Q_{yyy}\rangle+N_{yoo}\langle Q_{yoo}\rangle+N_{oyo}\langle Q_{oyo}\rangle\le
	\mathbb{E}^{U}(n_{yyy}+n_{yoo}+n_{oyo}).
\end{gather}

The joint constraints for 2 variables are:  
\begin{gather}
	N_{oyo}\langle Q_{oyo}\rangle+N_{ooy}\langle Q_{ooy}\rangle\le
	\mathbb{E}^{U}(n_{oyo}+n_{ooy}),\\
	N_{yoo}\langle Q_{yoo}\rangle+N_{ooy}\langle Q_{ooy}\rangle\le
	\mathbb{E}^{U}(n_{yoo}+n_{ooy}),\\
	N_{yoo}\langle Q_{yoo}\rangle+N_{oyo}\langle Q_{oyo}\rangle\le
	\mathbb{E}^{U}(n_{yoo}+n_{oyo}),\\
	N_{yyy}\langle Q_{yyy}\rangle+N_{ooy}\langle Q_{ooy}\rangle\le
	\mathbb{E}^{U}(n_{yyy}+n_{ooy}),\\
	N_{yyy}\langle Q_{yyy}\rangle+N_{oyo}\langle Q_{oyo}\rangle\le
	\mathbb{E}^{U}(n_{yyy}+n_{oyo}),\\
	N_{yyy}\langle Q_{yyy}\rangle+N_{yoo}\langle Q_{yoo}\rangle\le
	\mathbb{E}^{U}(n_{yyy}+n_{yoo}).
\end{gather}

\subsubsection{Range of $\mathcal{H}$}
\quad
In the case that the light sources of the three users are the same, and $N_{xxo}=N_{xox}=N_{oxx}$, $N_{xoo}=N_{oxo}=N_{oox}$, the bound of $\mathcal{H}$ can be expressed as:

\begin{equation}
	\mathcal{H}^L=e^{-\mu_x}\frac{\mathbb{E}^{L}(n_{oxx}+n_{xox}+n_{xxo})}{N_{oxx}}-e^{-2\mu_x}\frac{\mathbb{E}^{U}(n_{oox}+n_{oxo}+n_{xoo})}{N_{oox}}+e^{-3\mu_x}\frac{\mathbb{E}^{L}(n_{ooo})}{N_{ooo}},
\end{equation}
\begin{equation}
	\mathcal{H}^U=e^{-\mu_x}\frac{\mathbb{E}^{U}(n_{oxx}+n_{xox}+n_{xxo})}{N_{oxx}}-e^{-2\mu_x}\frac{\mathbb{E}^{L}(n_{oox}+n_{oxo}+n_{xoo})}{N_{oox}}+e^{-3\mu_x}\frac{\mathbb{E}^{U}(n_{ooo})}{N_{ooo}}.
\end{equation}

\subsubsection{Real value of $Y_{111}^{Z,L}$ and $e_{111}^{PZ,U}$}
To estimate the real value of the single-photon yield $Y_{111}^Z$ and the upper bound of the phase error rate $e_{111}^{PZ}$ of the Z-basis pulses, which cannot be directly observed in the experiment, we again use the Chernoff bound method to calculate the $Y_{111}^{Z,L}$ and $e_{111}^{PZ,U}$ according to the estimated expectation value $\langle Y_{111}^X\rangle^L$ and $\langle e_{111}^{BX}\rangle^U$:
\begin{equation}
    Y_{111}^{Z,L}=\frac{\mathbb{O}^L\left( N_{zzz}\mu_z^3e^{-3\mu_z}\langle Y_{111}^X\rangle^L, \epsilon \right)}{N_{zzz}\mu_z^3e^{-3\mu_z}},
\end{equation}
\begin{equation}
    e_{111}^{PZ,U}=\frac{\mathbb{O}^U\left( N_{zzz}\mu_z^3e^{-3\mu_z}Y_{111}^{Z,L}\langle e_{111}^{BX}\rangle^U, \epsilon \right)}{N_{zzz}\mu_z^3e^{-3\mu_z}Y_{111}^{Z,L}}.
\end{equation}
where $N_{zzz}$ is the number of the pulses prepared on the Z-basis, $\epsilon$ is the failure probability.

\subsubsection{Key rate with finite data size}
After calculating the range of $\mathcal{H}$, then the final secure key rate can be obtained by minimizing the $R(\mathcal{H})$ in the interval $[\mathcal{H}^L,\mathcal{H}^U]$:
\begin{equation}
    r=\mathrm{min}_{h\in [\mathcal{H}^L,\mathcal{H}^U]} R(h),
\end{equation}
which is also known as the single-scanning method. We remark that the double-scanning method \cite{4_intensity_qkd_2} also can be applied in the four-intensity MDI-QCC protocol.

\subsection{S3.3: Numerical Simulation}

\subsubsection{Comparison with infinite decoy-state method in the asymptotic case}
To analyze the security of the four-intensity MDI-QCC protocol, we first calculate the exact single-photon yield and the phase error rate of Z-basis photons with infinite decoy states, and compare its calculation results with the four-intensity method in the asymptotic case. 

Assuming that the users randomly encode Z-basis (X-basis) ideal single-photon states as $|H\rangle$ state or $|V\rangle$ state ($|+\rangle$ state or $|-\rangle$ state) with equal probability, the density matrix of Z-basis and X-basis pulses are:
	\begin{equation}
		\begin{split}
			\rho_Z&=\frac{1}{8}\left(|HHH\rangle\langle HHH|+|HHV\rangle\langle HHV|+|HVH\rangle\langle HVH|+|HVV\rangle\langle HVV|\right.\\&
			\quad\quad+\left.|VHH\rangle\langle VHH|+|VHV\rangle\langle VHV|+|VVH\rangle\langle VVH|+|VVV\rangle\langle VVV|\right)=\frac{I}{8},
		\end{split}
	\end{equation}
	
	\begin{equation}
		\begin{split}
			\rho_X&=\frac{1}{8}\left(|+++\rangle\langle +++|+|++-\rangle\langle ++-|+|+-+\rangle\langle +-+|+|+--\rangle\langle +--|\right.\\&
			\quad+\left.|-++\rangle\langle -++|+|-+-\rangle\langle -+-|+|--+\rangle\langle --+|+|---\rangle\langle ---|\right)=\frac{I}{8}.
		\end{split}
	\end{equation}
	The single-photon polarization state $|\psi_{111}\rangle$, prepared by three users, becomes a mixed state after passing through the loss channel with transmittance $\eta$:
	\begin{equation}
		\begin{split}
			|\psi_{111}\rangle\langle\psi_{111}|&\rightarrow \eta^3|\psi_{111}\rangle\langle\psi_{111}|+\eta^2(1-\eta)(|\psi_{110}\rangle\langle\psi_{110}|+|\psi_{101}\rangle\langle\psi_{101}|+|\psi_{011}\rangle\langle\psi_{011}|)\\&+\eta(1-\eta)^2(|\psi_{100}\rangle\langle\psi_{100}|+|\psi_{010}\rangle\langle\psi_{010}|+|\psi_{001}\rangle\langle\psi_{001}|)+(1-\eta)^3|\psi_{000}\rangle\langle\psi_{000}|,
		\end{split}
	\end{equation}
	where the subscripts denote the photon number for each senders. For clarity, we denote the density matrix above as:
	\begin{equation}
		\rho=\eta^3\rho_{3}+\eta^2(1-\eta)\rho_{2}+\eta(1-\eta)^2\rho_{1}+(1-\eta)^3\rho_{0},
	\end{equation}
	In the above formula, $\rho_{n}$ represents the density matrix of $n$ photon states. Then the yield of polarization state $|\psi\rangle$ is the sum of the yields of the different photon number states: 
	\begin{equation}
		Y^{|\psi\rangle}=\eta^3 Y^{|\psi\rangle}_{111}+\eta^2(1-\eta)\left(Y^{|\psi\rangle}_{011}+Y^{|\psi\rangle}_{101}+Y^{|\psi\rangle}_{110}\right)+\eta(1-\eta)^2\left(Y^{|\psi\rangle}_{100}+Y^{|\psi\rangle}_{010}+Y^{|\psi\rangle}_{001}\right)+(1-\eta)^3Y^{|\psi\rangle}_{000}.
	\end{equation}
	And the yields of Z-basis and X-basis pulses are:
	\begin{equation}
		Y_Z=\frac{1}{8}\left(Y^{|HHH\rangle}+Y^{|HHV\rangle}+Y^{|HVH\rangle}+Y^{|HVV\rangle}+Y^{|VHH\rangle}+Y^{|VHV\rangle}+Y^{|VVH\rangle}+Y^{|VVV\rangle}\right),
	\end{equation}
	\begin{equation}
		Y_X=\frac{1}{8}\left(Y^{|+++\rangle}+Y^{|++-\rangle}+Y^{|+-+\rangle}+Y^{|+--\rangle}+Y^{|-++\rangle}+Y^{|-+-\rangle}+Y^{|--+\rangle}+Y^{|---\rangle}\right).
	\end{equation}
	
	In the GHZ-state analyzer shown in the Figure 1, we assume that the dark count rate of detectors is $p_d$. If only one detector of each path clicks, then such a 3-fold coincidence count corresponds to a successful projection measurement. Considering the dark count rate $p_d$ of the detector, the probability that only one detector of a path clicks is $1-p_d$ for the single photon entering into that path and $2p_d(1-p_d)$ for the vacuum state: 
     \begin{equation}
         p_{\text{vacuum state}}=1-p_d^2-(1-p_d)^2=2p_d(1-p_d)
     \end{equation}
 For states with different photon number and polarization input to the GHZ-state analyzer, we calculate their evolution and yield as follows respectively: 
	
	\begin{itemize}
		\item[(i)]Z-basis: 
		
		\textbf{Yield of $\bm{\rho_3}$}. If all single photons pass through the channel, the evolution of the polarization state in the GHZ-state analyzer is:
		
		\begin{equation}
			\begin{split}
				&|H\rangle_A |H\rangle_B |H\rangle_C\rightarrow |H\rangle_1 |H\rangle_2 |H\rangle_3, \quad
				|V\rangle_A |V\rangle_B |V\rangle_C\rightarrow |V\rangle_1 |V\rangle_2 |V\rangle_3,\\&
				|H\rangle_A |H\rangle_B |V\rangle_C\rightarrow |H\rangle_1 |H\rangle_3 |V\rangle_3, \quad
				|V\rangle_A |V\rangle_B |H\rangle_C\rightarrow |V\rangle_1 |H\rangle_2 |V\rangle_2,\\&
				|H\rangle_A |V\rangle_B |H\rangle_C\rightarrow |H\rangle_3 |H\rangle_2 |V\rangle_2, \quad
				|V\rangle_A |H\rangle_B |V\rangle_C\rightarrow |H\rangle_1 |V\rangle_1 |V\rangle_3,\\&
				|V\rangle_A |H\rangle_B |H\rangle_C\rightarrow |H\rangle_1 |V\rangle_1 |H\rangle_2, \quad
				|H\rangle_A |V\rangle_B |V\rangle_C\rightarrow |V\rangle_2 |H\rangle_2 |V\rangle_3,
			\end{split}
		\end{equation}
	where the subscripts $A$, $B$, $C$ on left hand side represent users and the subscripts $1$, $2$, $3$ on right hand side represent paths. For the first two states above, three photons are input into different paths and their yields are:
	
		\begin{equation}
    Y^{|HHH\rangle}_{111}=Y^{|VVV\rangle}_{111}=(1-p_d)^3.
		\end{equation}
	
	For the latter six states, two photons on the same path are incident on the same detector due to the HOM interference effect after passing sequentially through the HWP with optical axis of $22.5^{\circ}$ and PBS. Thus, the 3-fold coincidence count will be obtained only when the dark count causes the detector of the path without a photon to click. Their yields are:
	
	 \begin{equation}
Y^{|VHH\rangle}_{111}=Y^{|HVH\rangle}_{111}=Y^{|HHV\rangle}_{111}=Y^{|HVV\rangle}_{111}=Y^{|VHV\rangle}_{111}=Y^{|VVH\rangle}_{111}=2(1-p_d)^3p_d.
	 \end{equation}
	
	\textbf{Yield of $\bm{\rho_2}$}. If one of the three photons is dissipated and the remaining two photons are in the different path after passing through the PBS, the yields in this case are:
	 
		\begin{equation}
Y^{|HHH\rangle}_{110}=Y^{|HHH\rangle}_{101}=Y^{|HHH\rangle}_{011}=Y^{|VVV\rangle}_{110}=Y^{|VVV\rangle}_{101}=Y^{|VVV\rangle}_{011}=2(1-p_d)^3p_d,
		\end{equation}
		
		\begin{equation}
Y^{|HHV\rangle}_{110}=Y^{|HVH\rangle}_{101}=Y^{|VHH\rangle}_{011}=Y^{|VVH\rangle}_{110}=Y^{|VHV\rangle}_{101}=Y^{|HVV\rangle}_{011}=2(1-p_d)^3p_d,
		\end{equation}
		
		\begin{equation}
Y^{|HVV\rangle}_{110}=Y^{|HVH\rangle}_{110}=Y^{|VHH\rangle}_{101}=Y^{|VVH\rangle}_{101}=Y^{|HHV\rangle}_{011}=Y^{|VHV\rangle}_{011}=2(1-p_d)^3p_d.
		\end{equation}
		
	Besides, if the remaining two photons are on the same path after passing through the PBS, the yields are:
	
		\begin{equation}
Y^{|VHH\rangle}_{110}=Y^{|VHV\rangle}_{110}=Y^{|HHV\rangle}_{101}=Y^{|HVV\rangle}_{101}=Y^{|VVH\rangle}_{011}=Y^{|HVH\rangle}_{011}=4(1-p_d)^3p_d^2.
		\end{equation}
	
	\textbf{Yield of $\bm{\rho_1}$}. When two of three photons are dissipated, the yields for all Z-basis states are:
	
		\begin{equation}
Y^{|\psi\rangle}_{100}=Y^{|\psi\rangle}_{010}=Y^{|\psi\rangle}_{001}=4(1-p_d)^3p_d^2
		\end{equation}
	
		\item[(ii)] X-basis: 
		
		\textbf{Yield of $\bm{\rho_3}$}. We assume that users sent single photon with polarization state $\left(|H\rangle+e^{i\theta_{A,B,C}}|V\rangle\right)/\sqrt{2}$, where $\theta_{A,B,C}=0$ or $\pi$. In the case of all single photons passing through the channel, the polarization state in the analyzer evolves as follows:
		\begin{equation}
			\begin{aligned}
				\frac{1}{2\sqrt{2}}\left(|H\rangle_A+e^{i\theta_A}|V\rangle_A\right)\left(|H\rangle_B+e^{i\theta_B}|V\rangle_B\right)\left(|H\rangle_C+e^{i\theta_C}|V\rangle_C\right)  \stackrel{PBSs}{\longrightarrow} \\ \frac{1}{2\sqrt{2}}\left(|H\rangle_1|H\rangle_2|H\rangle_3+e^{i(\theta_A+\theta_B+\theta_C)}|V\rangle_1|V\rangle_2|V\rangle_3+e^{i\theta_A}|H\rangle_1|V\rangle_1|H\rangle_2+e^{i(\theta_A+\theta_C)}|H\rangle_1|V\rangle_1|V\rangle_3+\right. \\ \left. e^{i\theta_B}|H\rangle_2|V\rangle_2|H\rangle_3+e^{i(\theta_A+\theta_B)}|V\rangle_1|H\rangle_2|V\rangle_2+e^{i\theta_C}|H\rangle_1|H\rangle_3|V\rangle_3+e^{i(\theta_B+\theta_C)}|V\rangle_2|H\rangle_3|V\rangle_3\right).
			\end{aligned}
		\end{equation}
		The probability of obtaining an effective 3-fold coincidence count for the first two components of the state after PBSs is $(1-p_d)^3$ and $2p_d(1-p_d)^3$ for the latter six components based on the results of previous section. Thus, the yields of X-basis photons without loss are:
		\begin{equation}
			Y^{|\pm\pm\pm\rangle}_{111}=\frac{1}{4}(1-p_d)^3+\frac{3}{2}(1-p_d)^3p_d.
		\end{equation}
	
		\textbf{Yield of $\bm{\rho_2}$}. The evolution of the state when one of the three photons is dissipated can be calculated as:	
		\begin{equation}
			\begin{split}
				&\frac{1}{2}\left(|H\rangle_A+e^{i\theta_A}|V\rangle_A\right)\left(|H\rangle_B+e^{i\theta_B}|V\rangle_B\right)  \stackrel{PBSs}{\longrightarrow}\\&  \frac{1}{2}\left(|H\rangle_1|H\rangle_3+e^{i\theta_A}|V\rangle_1|H\rangle_1+e^{i\theta_B}|V\rangle_2|H\rangle_3+e^{i(\theta_A+\theta_B)}|V\rangle_1|V\rangle_2 \right),
			\end{split}
		\end{equation}

		\begin{equation}
			\begin{split}
				&\frac{1}{2}\left(|H\rangle_A+e^{i\theta_A}|V\rangle_A\right)\left(|H\rangle_C+e^{i\theta_C}|V\rangle_C\right)  \stackrel{PBSs}{\longrightarrow}\\&  \frac{1}{2}\left(|H\rangle_2|H\rangle_3+e^{i\theta_A}|V\rangle_1|H\rangle_2+e^{i\theta_C}|V\rangle_3|H\rangle_3+e^{i(\theta_A+\theta_C)}|V\rangle_1|V\rangle_3 \right),
			\end{split}
		\end{equation}
	
		\begin{equation}
			\begin{split}
				&\frac{1}{2}\left(|H\rangle_B+e^{i\theta_B}|V\rangle_B\right)\left(|H\rangle_C+e^{i\theta_C}|V\rangle_C\right)  \stackrel{PBSs}{\longrightarrow}\\&  \frac{1}{2}\left(|H\rangle_1|H\rangle_2+e^{i\theta_B}|V\rangle_2|H\rangle_2+e^{i\theta_C}|V\rangle_3|H\rangle_1+e^{i(\theta_B+\theta_C)}|V\rangle_2|V\rangle_3 \right).
			\end{split}
		\end{equation}
		Similarly, we have
		\begin{equation}
			Y^{|+++\rangle}_{110}=Y^{|+++\rangle}_{101}=Y^{|+++\rangle}_{011}=Y^{|---\rangle}_{110}=Y^{|---\rangle}_{101}=Y^{|---\rangle}_{011}=\frac{3}{2}(1-p_d)^3p_d+p_d^2(1-p_d)^3,
		\end{equation}
	
		\begin{equation}
			Y^{|++-\rangle}_{110}=Y^{|+-+\rangle}_{101}=Y^{|-++\rangle}_{011}=Y^{|--+\rangle}_{110}=Y^{|-+-\rangle}_{101}=Y^{|+--\rangle}_{011}=\frac{3}{2}(1-p_d)^3p_d+p_d^2(1-p_d)^3,
		\end{equation}
	
		\begin{equation}
			Y^{|+-+\rangle}_{110}=Y^{|+--\rangle}_{110}=Y^{|++-\rangle}_{101}=Y^{|+--\rangle}_{101}=Y^{|++-\rangle}_{011}=Y^{|-+-\rangle}_{011}=\frac{3}{2}(1-p_d)^3p_d+p_d^2(1-p_d)^3,
		\end{equation}
	
		\begin{equation}
			Y^{|-+-\rangle}_{110}=Y^{|-++\rangle}_{110}=Y^{|--+\rangle}_{101}=Y^{|-++\rangle}_{101}=Y^{|--+\rangle}_{011}=Y^{|+-+\rangle}_{011}=\frac{3}{2}(1-p_d)^3p_d+p_d^2(1-p_d)^3.
		\end{equation}
	
	
		\textbf{Yield of $\bm{\rho_1}$}. When two of the three photons are dissipated, the yields for all X-basis states are:
		\begin{equation}
			Y^{|\psi\rangle}_{100}=Y^{|\psi\rangle}_{010}=Y^{|\psi\rangle}_{001}=4(1-p_d)^3p_d^2.
		\end{equation}
	
	\item[(iii)] \textbf{Yield of $\bm{\rho_0}$}. For all states, the yield of the vacuum state is:
	\begin{equation}
		Y_{000}=8p_d^3(1-p_d)^3.
	\end{equation}
	\end{itemize}

    From the results above, the yields of different single-photon polarization states of Z-basis are:
	
	\begin{equation}
		Y_{HHH}=Y_{VVV}=(1-p_d)^3\left[\eta^3+6\eta^2(1-\eta)p_d+12\eta(1-\eta)^2p_d^2+8(1-\eta)^3p_d^3\right],
	\end{equation}
	
	\begin{equation}
		\begin{split}
			&Y_{HHV,HVH,HVV,VVH,VHV,VHH}=(1-p_d)^3\left[2\eta^3p_d+4\eta^2(1-\eta)(p_d+p_d^2)+12\eta(1-\eta)^2p_d^2+8(1-\eta)^3p_d^3 \right].
		\end{split}
	\end{equation}
	For the polarization states encoded on X-basis, all yields are the same:
	\begin{equation}
		\begin{split}
			Y_{\pm\pm\pm}=(1-p_d)^3\left[\frac{1}{4}\eta^3+\frac{3}{2}\eta^3p_d+\frac{9}{2}\eta^2(1-\eta)p_d+3\eta^2(1-\eta)p_d^2+12\eta(1-\eta)^2p_d^2+8(1-\eta)^3p_d^3 \right].
		\end{split}
	\end{equation}
	
	Hence the overall single-photon yield of the Z-basis and X-basis polarization states are:
	\begin{equation}
		Y_{111}^Z=(1-p_d)^3\left[\frac{1}{4}\eta^3+\frac{3}{2}\eta^3p_d+\frac{9}{2}\eta^2(1-\eta)p_d+3\eta^2(1-\eta)p_d^2+12\eta(1-\eta)^2p_d^2+8(1-\eta)^3p_d^3 \right],
	\end{equation}
	
	\begin{equation}
		Y_{111}^X=(1-p_d)^3\left[\frac{1}{4}\eta^3+\frac{3}{2}\eta^3p_d+\frac{9}{2}\eta^2(1-\eta)p_d+3\eta^2(1-\eta)p_d^2+12\eta(1-\eta)^2p_d^2+8(1-\eta)^3p_d^3 \right].
	\end{equation}
	It can be concluded that $Y_{111}^Z=Y_{111}^X=Y_{111}$,i.e. the yields of single-photon pulses on Z-basis and X-basis are equal in the asymptotic case. On the other hand, assuming that the misalignment error is $e_d$, and considering that the vacuum state imposes a error rate of 0.5, then the single-photon bit error rate of the X-basis is:
	\begin{equation}
		e^{BX}_{111}=(1-p_d)^3\left[\frac{e_d}{4}\eta^3+\frac{3}{4}\eta^3p_d+\frac{9}{4}\eta^2(1-\eta)p_d+\frac{3}{2}\eta^2(1-\eta)p_d^2+6\eta(1-\eta)^2p_d^2+4(1-\eta)^3p_d^3 \right]/Y_{111}.
	\end{equation}
	 According to the conclusion that $e_{111}^{PZ}=e_{111}^{BX}$, we can calculate the phase error rate of Z-basis based on the formula above.
 
	We calculate the $Y_{111}^Z$, $e_{111}^{PZ}$ and key rate with asymptotic formula assuming infinite decoy-state protocol derived above and four-intensity method, respectively. We show the result of using four-intensity method (black), which gives a key rate that is nearly the same as the corresponding one using the infinite decoy states (red) in the figures below:
	\begin{figure}[H]
		\centering
		\includegraphics[width=\textwidth]{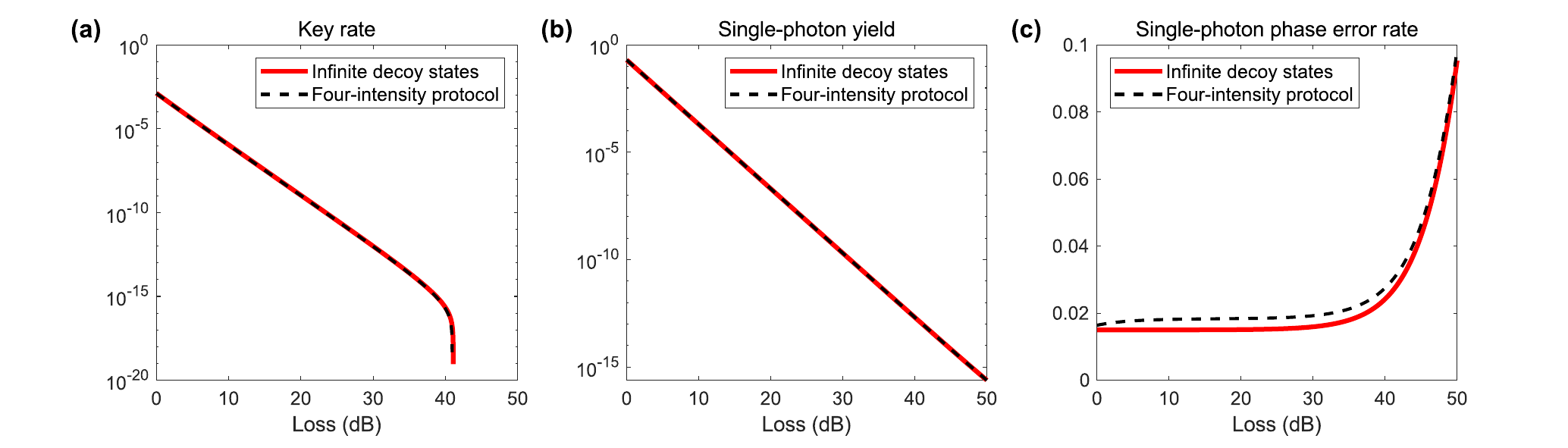}
		\caption{Comparison of calculation results of infinite decoy-state method and four-intensity method. (a) key rate; (b) single-photon yield; (c) single-photon phase error rate. The parameters used in the calculation are: detection efficiency $\eta_d=0.93$, dark count rate $p_d=10^{-7}$, misalignment error $e_d=1.5\%$.}
	\end{figure}

\subsubsection{Comparison with three-intensity decoy-state method in non-asymptotic case}
\quad
We simulate the conference key rate for the three-intensity and four-intensity protocol under different conditions. The parameters used in the simulation are shown in the table below:

\begin{table}[H]
\centering
\begin{tabular}{@{}cccccc@{}}
\toprule
$p_d$   & $\eta_d$ & $e_d$   & $f$     & $\varepsilon$ & $V$\\ \midrule
$10^{-6}$ & $0.8$  & $0.025$ & $1.16$ & $10^{-10}$ & $0.25$       \\ \bottomrule
\end{tabular}
\caption{Parameters in simulation}
\end{table}

The secure key rate formula of three-intensity protocol we use in simulation is:
\begin{equation}
    R_{\text{three-intensity}}=(p_{Z|\mu}p_{\mu})^3\left\{\mu^3e^{-3\mu}{Y}_{111}^{Z,L}\left[1-H({e}_{111}^{PZ,U}) \right]-Q_{\mu\mu\mu}^ZfH(E^Z_{\mu\mu\mu}) \right\},
\end{equation}
where $\mu$ is the average photon number of the signal state pulses, $p_{\mu}$ is the probability of selecting the signal source, $p_{Z|\mu}$ is the probability of encoding pulses on Z-basis under the condition of selecting the signal source, ${Y}_{111}^{Z,L}$ is the lower bound of the yield of Z-basis single-photon component, ${e}_{111}^{PZ,U}$ is the upper bound of the phase error rate of Z-basis single-photon component; $Q_{\mu\mu\mu}^Z$ and $E^Z_{\mu\mu\mu}$ is the overall gain and the quantum bit error rate of the Z-basis pulses from the signal source respectively. In the numerical simulation of the three-intensity protocol, the key rate is the function of six parameters: $\mu$, $\nu$, $p_{\mu}$, $p_{\nu}$, $p_{Z|\mu}$ and $p_{Z|\nu}$. We implement the full parameter optimization.

We first compare the optimal key rates of the two protocols under different total channel losses for $N=10^{13}$ and $N=10^{14}$, respectively. The results show that the key rate and the communication distance of the four-intensity protocol is tremendously improved over the three-intensity protocol in both cases. In particular, there is a one- to two-order-of-magnitude improvement in the conference key rate when the number of pulses is small and the total channel loss is large. On the other hand, we perform key rate optimization by varying the number of pulses with a total channel loss of 18 dB. The results show that the four-intensity protocol can generate the secure key by sending only $3.5\times10^{12}$ pulses, however, the three-intensity protocol requires a minimum of about $4\times10^{13}$. This implies that the four-intensity protocol has a lower threshold for the number of pulses required for generating key and is therefore experimentally easier to implement. 

\begin{figure}[H]
    \centering
    \includegraphics[width=1\linewidth]{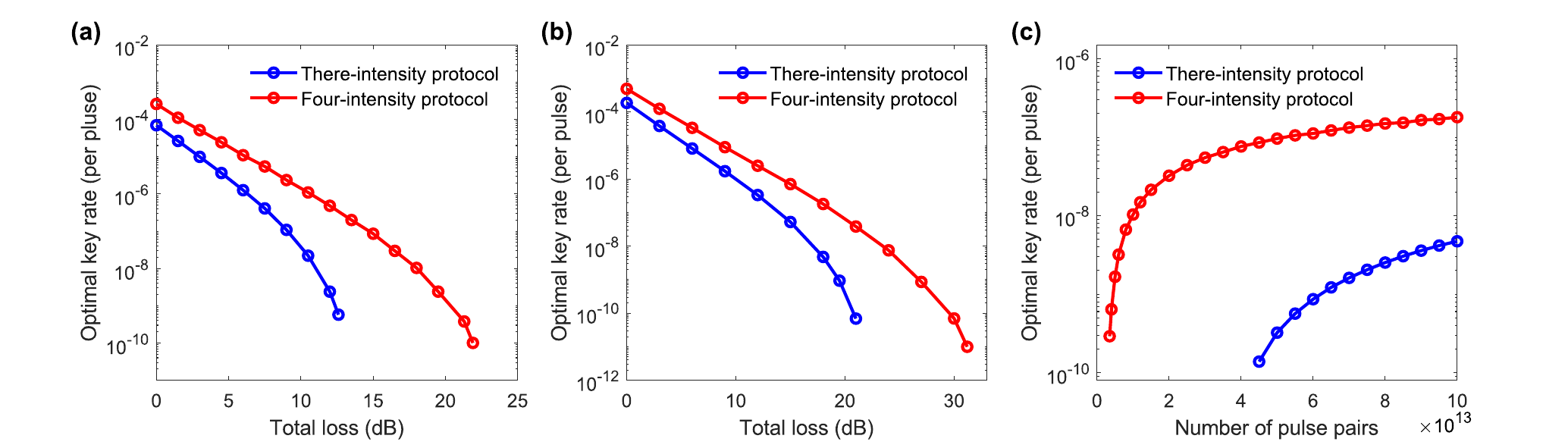}
    \caption{{Optimal conference key rate of three-intensity protocol and four-intensity protocol}. (a) $N=10^{13}$. (b) $N=10^{14}$. (c) Optimal key rate at a total loss of 18 dB with different pulse numbers sent by users.}
    \label{fig:theory}
\end{figure}

\begin{table}[H]
\centering
\begin{tabular}{@{}ccccc@{}}
\toprule
                                                       & \multicolumn{2}{c}{$N=10^{13}$}             & \multicolumn{2}{c}{$N=10^{14}$}             \\ \midrule
Total loss (dB)                                        & $6.0$                & $12.0$               & $9.0$                & $18.0$               \\
$R_{\text{3-int}}$                           & $1.267\times10^{-6}$ & $2.385\times10^{-9}$ & $1.720\times10^{-6}$ & $4.764\times10^{-9}$ \\
$R_{\text{4-int}}$                            & $1.101\times10^{-5}$ & $4.820\times10^{-7}$ & $8.949\times10^{-6}$ & $1.804\times10^{-7}$ \\
$R_{\text{4-int}}/R_{\text{3-int}}$ & $8.69$              & $202.10$             & $5.20$               & $37.87$              \\ \bottomrule
\end{tabular}
\caption{Comparison of conference key rate of two protocols for different sent pulse number and channel loss.}
\end{table}

\section{S4: Experimental results}

\begin{table}[H]
\setlength{\tabcolsep}{22pt}
\renewcommand\arraystretch{1.5}
\centering
\scalebox{0.75}{
\begin{tabular}{|c|c|ccc|}
\hline\hline
\textbf{Overall attenuation (dB)}    & $\eta$    & 14.1    & 17.8    & 21.5\\ \hline
\multirow{8}{*}{\begin{tabular}[c]{@{}c@{}}\textbf{Number of} \\ \textbf{pulse pairs}\end{tabular}}  & $N_{zzz}$                 & \multicolumn{3}{c|}{$7.2000\times 10^{11}$} \\
                                & $N_{xxx}$                 & \multicolumn{3}{c|}{$2.6528\times 10^{12}$} \\
                                & $N_{yyy}$                 & \multicolumn{3}{c|}{$1.6000\times 10^{10}$} \\
                                & $N_{ooo}$                 & \multicolumn{3}{c|}{$6.0000\times 10^{9}$}   \\
                                & $N_{oxx}+N_{xox}+N_{xxo}$ & \multicolumn{3}{c|}{$1.1040\times 10^{12}$} \\
                                & $N_{xoo}+N_{oxo}+N_{oox}$ & \multicolumn{3}{c|}{$1.4400\times 10^{11}$} \\
                                & $N_{oyy}+N_{yoy}+N_{yyo}$ & \multicolumn{3}{c|}{$2.4000\times 10^{10}$} \\
                                & $N_{yoo}+N_{oyo}+N_{ooy}$ & \multicolumn{3}{c|}{$2.4000\times 10^{10}$} \\ \hline
\multirow{12}{*}{\textbf{Total gains}} & $M_{zzz}$                 & $6851252$     & $3061962$    & $1248883$    \\
    & $M_{xxx}$                 & $2207217$     & $951388$     & $408827$     \\
    & $M_{yyy}$                 & $1941154$     & $857992$     & $373106$     \\
    & $M_{oyy}$                 & $233950$      & $105880$      & $44065$      \\
    & $M_{yoy}$                 & $280433$      & $114446$      & $49702$      \\
    & $M_{yyo}$                 & $261515$      & $118246$      & $49535$      \\
    & $M_{yoo}$                 & $176$          & $60$         & $29$         \\
    & $M_{oyo}$                 & $5$           & $3$          & $0$          \\
    & $M_{ooy}$                 & $380$         & $115$         & $49$         \\
    & $M_{ooo}$                 & $0$           & $0$          & $0$          \\
    & $M_{oxx}+M_{xox}+M_{xxo}$ & $237638$      & $101573$      & $42814$      \\
    & $M_{xoo}+M_{oxo}+M_{oox}$ & $35$          & $14$          & $6$          \\ \hline
\multirow{5}{*}{\textbf{Error gains}}& $E_{zzz}^{AB}$                & $131768$      & $68298$      & $22716$      \\
                                & $E_{zzz}^{AC}$                & $132496$      & $71445$      & $23973$      \\
                                & $E_{zzz}^{BC}$                & $150078$      & $73779$      & $26937$      \\
                                & $E_{xxx}$                 & $882539$      & $379058$     & $162357$     \\
                                & $E_{yyy}$                 & $831142$      & $368415$     & $158805$     \\ \hline
\multirow{7}{*}{\textbf{Results}}    & $e_Z^{AB}$ (\%)   & $1.92$    & $2.23$    & $1.82$\\
                                & $e_Z^{AC}$ (\%)   & $1.93$    & $2.33$    & $1.92$\\
                                & $e_Z^{BC}$ (\%)   & $2.19$    & $2.41$    & $2.16$\\
                                & $e_{111}^{PZ}$ (\%)  & $15.04$    & $17.63$    & $22.16$\\
                                & $Y_{111}^{Z}$    & $8.70\times 10^{-3}$    & $3.86\times 10^{-3}$    & $1.78\times 10^{-3}$\\
                                & key rate (per pulse)    & $3.02\times 10^{-8}$    & $4.69\times 10^{-9}$    & $3.86\times 10^{-10}$ \\
                                \hline\hline
\end{tabular}
}
\caption{Detailed results of the MDI-QCC experiment. The subscripts indicate the photon sources of three users. We remark that the SNSPDs have the detection efficiency of about 80\% and the dark count rate of about $10^{-6}$. To reduce the error rate, we set the coincidence time window to 156.25 ps in the data processing.
The efficiency of the coincidence time window is about 87\%, 90\%, 91\% for the data of overall attenuation of 14.1 dB, 17.8 dB and 21.5 dB respectively.}
\end{table}

\nocite{*}
\bibliography{mdi_qcc_bib.bib}

\end{document}